\documentclass[review,12pt]{elsarticle}
\usepackage{newtxtext,newtxmath}

\biboptions{square,comma,sort&compress}

\usepackage{amssymb} 
\usepackage{amsthm} 
\usepackage{amsmath}
\usepackage{mathrsfs}
\usepackage{graphicx}
\usepackage{epstopdf}
\usepackage{float}
\usepackage{caption}
\usepackage{soul}
\usepackage{color,soul} 
\usepackage{color} 
\usepackage{subcaption}
\usepackage{bm}
\usepackage{bbm}
\usepackage{mathrsfs}
\usepackage{soul}
\usepackage[margin=2cm]{geometry}
\usepackage{booktabs}
\usepackage{chngpage}
\usepackage{enumitem}
\usepackage{subcaption}
\usepackage{multirow}
\usepackage{longtable}
\usepackage{makecell}
\usepackage{stackengine} 
\usepackage{mhchem} 
\usepackage{hyperref} 
\usepackage{cleveref}
\usepackage{makecell} 
\usepackage{tikz}
\usetikzlibrary{decorations.markings, arrows.meta}
\usepackage{tikz-3dplot}
\usetikzlibrary{
arrows,
arrows.meta,
automata,
backgrounds,
calc,
decorations,
decorations.pathmorphing,
decorations.pathreplacing,
decorations.fractals,
external,
fit,
matrix,
petri,
positioning,
shadows,
shapes,
shapes.multipart,
topaths,
intersections
}

\usepackage{array} 
\journal{Experimental mechanics}

\makeatletter
\def\@author#1{\g@addto@macro\elsauthors{\normalsize%
    \def\baselinestretch{1}%
    \upshape\authorsep#1\unskip\textsuperscript{%
      \ifx\@fnmark\@empty\else\unskip\sep\@fnmark\let\sep=,\fi
      \ifx\@corref\@empty\else\unskip\sep\@corref\let\sep=,\fi
      }%
    \def\authorsep{\unskip,\space}%
    \global\let\@fnmark\@empty
    \global\let\@corref\@empty  
    \global\let\sep\@empty}%
    \@eadauthor={#1}
}
\makeatother

\makeatletter
\def\thickhline{%
  \noalign{\ifnum0=`}\fi\hrule \@height \thickarrayrulewidth \futurelet
   \reserved@a\@xthickhline}
\def\@xthickhline{\ifx\reserved@a\thickhline
               \vskip\doublerulesep
               \vskip-\thickarrayrulewidth
             \fi
      \ifnum0=`{\fi}}
\makeatother

\newlength{\thickarrayrulewidth}
\begin{document}

\begin{frontmatter}



\title{High-sensitivity hydrogen gas permeation: system development, sample preparation, and influence of testing variables}


\author{Rongrui Li\fnref{OX}}

\author{Alfredo Zafra\fnref{OX}}

\author{Zachary D. Harris\fnref{Pitt}}

\author{Emilio Mart\'{\i}nez-Pañeda\corref{cor1}\fnref{OX}}
\ead{emilio.martinez-paneda@eng.ox.ac.uk}

\address[OX]{Department of Engineering Science, University of Oxford, Oxford OX1 3PJ, UK}

\address[Pitt]{Mechanical Engineering and Materials Science Department, University of Pittsburgh, Pittsburgh, PA 15261, USA}

\cortext[cor1]{Corresponding author.}

\begin{abstract}
\textbf{Background} There is a need to develop quantitative, high-resolution hydrogen gas (H$_2$) permeation techniques to provide a better understanding of hydrogen-material interactions, from hydrogen uptake and diffusion to embrittlement.\vspace{8pt}

\noindent \textbf{Objective} This study aims to develop and validate a high-sensitivity H$_2$ permeation system, which is then leveraged to systematically quantify the influence of surface condition and key testing variables (surface oxides, residual gas impurities, pressure, and temperature) on hydrogen permeation. \vspace{8pt}

\noindent \textbf{Methods} A gas permeation system capable of operating at pressures up to 50 bar and temperatures up to 250 $^\circ$C was developed, incorporating high-sensitivity mass spectrometric detection and controlled surface preparation protocols. Permeation transients were analysed in a model material (annealed pure Fe) to determine hydrogen diffusivity and permeability under systematically varied surface states, oxygen contents, pressures and temperatures. \vspace{8pt}

\noindent \textbf{Results} 
Surface oxides are shown to play a dominant role in controlling hydrogen permeation at room temperature. The presence of oxide layers can severely hinder or completely suppress hydrogen uptake, with measurable permeation requiring oxide removal via pickling and Pd coating on both surfaces, or activation through hydrogen-assisted reduction at elevated temperature. Residual oxygen present prior to hydrogen exposure further reduces permeability by modifying surface boundary conditions, indicating strongly surface-controlled kinetics. Under optimised surface conditions, hydrogen transport follows bulk diffusion-controlled behaviour, with steady-state flux obeying Sieverts’ law at 25 $^\circ$C (1-5 bar) and diffusivity and permeability exhibiting Arrhenius behaviour between 25 and 150 $^\circ$C at 5~bar, indicating bulk diffusion-controlled transport. \vspace{8pt}

\noindent \textbf{Conclusions} The developed high-sensitivity permeation system resolves hydrogen fluxes as low as 1.98 $\times$ 10$^{-9}$ mol/(m$^2$$\cdot$s) and provides a robust platform for investigating surface, mechanical and environmental effects on hydrogen permeation under realistic service conditions.

\vspace{1.0em}

\end{abstract}
\end{frontmatter}

\section{Introduction}

Hydrogen is central to decarbonising hard-to-electrify sectors and enabling long-term, large-scale energy storage. Therefore, the use of hydrogen -- already prominent in processes such as refining, ammonia production and steelmaking -- is expected to significantly broaden over the next decade \cite{gunathilake_comprehensive_2024}. As a result, a vast number of structural components, containment systems (e.g. pressure vessels, storage tanks), and pipelines will be in contact with hydrogen, typically in its gaseous form (H$_2$). However, the deployment of a large-scale hydrogen infrastructure is  challenged by the phenomenon of \emph{hydrogen embrittlement}, whereby metals exposed to H$_2$ experience a significant reduction in ductility, fracture toughness and fatigue crack growth resistance \cite{chen_hydrogen_2025,delrio2025local,malheiros2022local,fernandez-sousa_analysis_2020}.\\ 

Hydrogen ingress and transport kinetics, together with bulk trapping mechanisms, control the local hydrogen supply available to participate in fracture processes proximate to a crack tip, thereby having a critical effect on fracture behaviour \cite{martinez2016strain,guedes2020role}. As such, preventing and mitigating hydrogen embrittlement requires accurate quantification of hydrogen availability in susceptible regions of the microstructure under relevant environments. Hydrogen gas permeation is one of the most suitable techniques for this purpose, as it provides a direct and quantitative link between a given hydrogen-containing environment and the resulting hydrogen flux through the metal. By analysing transient permeation responses under controlled boundary conditions, hydrogen-metal interaction parameters such as permeability, solubility and diffusivity can be obtained, which are necessary for engineering design and the development of predictive models. However, gas permeation data under service-relevant operating conditions for H$_2$ infrastructure remain scarce, as the technique is experimentally complex, involves safety constraints intrinsic to operating with gaseous hydrogen, has not been widely implemented across laboratories, and lacks standardisation and systematic understanding of how testing variables influence the measured response.\\

Alternatively, electrochemical permeation has historically been the most widely used method, due to its standardisation (ISO 17081, ASTM G148) \cite{noauthor_iso_2014, noauthor_standard_2018}, relatively low cost, and ease of implementation. However, it is not suited to investigate H$_2$-specific surface phenomena as measurements are susceptible to electrochemistry-related artefacts such as evolving surface conditions, bubble formation, and pH/solution composition changes \cite{cupertino2024suitability,bao2026new}, which hinder reproducibility \cite{zafra_comparison_2022}. In addition, the evolution and stability of surface oxides in an electrochemical environment differs significantly from that of the hydrogen gas counterpart. Moreover, the use of electrochemical permeation is generally limited to relatively low temperatures ($\textless$ 80 $^\circ$C) due to electrolyte stability. Despite these limitations, attempts have been made to relate electrochemical and gas permeation through reaction rate approaches \cite{hageman2022electro,cupertino2024hydrogen} or \textit{via} the concept of equivalent hydrogen fugacity \cite{liu_determination_2014,koren_experimental_2023}; however, extrapolation to real gas environments remains elusive and largely confined to research applications. Hybrid gaseous-electrochemical approaches have also been proposed, for example by combining a hydrogen pressure charging side with an electrochemical detection side \cite{wang_study_2023, koren_experimental_2023, gao_synergic_2024}. Nevertheless, electrochemical detection remains constrained by the aforementioned limitations, particularly temperature restrictions and the indirect nature of the measurement, which relies on hydrogen oxidation currents rather than direct flux quantification. In addition, unlike electrochemical permeation, gaseous permeation systems can be used to characterise hydrogen uptake and transport in non-electrically conductive materials, such as polymers.\\

Consequently, there is a growing need for fully gaseous hydrogen permeation systems that enable measurements using pure hydrogen or controlled gas mixtures, avoid electrochemical artefacts, replicate realistic service conditions, and allow direct correlation between applied gas pressure, hydrogen flux, and diffusivity over a wide range of temperatures. This is particularly important for low-diffusivity materials and complex systems such as coated steels or bimetallic architectures. However, despite this need, there is currently no standardised methodology for gaseous hydrogen permeation testing of metals, and existing systems differ significantly in design, boundary conditions, detection strategies, and surface preparation procedures \cite{fan_hydrogen_2025,okayasu2021examination}. Accurate interpretation of gaseous permeation data requires consideration of the individual physical processes governing hydrogen transport, including H$_2$ adsorption and dissociation at the surface, entry across surface films (e.g., oxides or coatings), diffusion through the bulk microstructure, and recombination/desorption at the exit surface \cite{protopopoff2002surface,zhang_gaseous_2025}. As these processes also occur during hydrogen uptake in real components, their relative contributions must be carefully controlled and understood. However, isolating and quantifying these individual contributions remains challenging due to the lack of well-defined experimental protocols. Recent work has highlighted the need for improved standardisation of gaseous permeation methodologies, particularly with respect to apparatus calibration, system conditioning, and specimen surface control \cite{fan_hydrogen_2025,zhang_gaseous_2025}. In addition, variability in detection techniques further complicates data interpretation and comparison. Traditional pressure-rise methods are relatively easier to set up but lack sensitivity at low hydrogen fluxes (e.g., those associated with room temperature conditions), whereas gas chromatography (GC) and mass spectrometry (MS) offer improved sensitivity and selectivity at the expense of increased experimental complexity, cost, and calibration requirements \cite{rothig_gaseous_2025,fujiwara_high-pressure_2020,huang_effect_2020,sand_versatile_2024}.\\

Among the experimental factors that remain poorly understood, two phenomena are of particular interest from both fundamental research and engineering practice perspectives - how hydrogen permeation is affected by: (i) surface oxides, and (ii) gas impurities. For example, a native oxide layer readily forms on iron and steel surfaces under ambient conditions. Although only a few nanometers thick, this oxide layer has been reported to significantly impact hydrogen uptake by altering surface-controlled processes such as adsorption, dissociation, and recombination \cite{wampler_surface-limited_1986,wampler_hydrogen_1989,AnodicIronOxide1992a,roosendaal_passivation_1999,grosvenor_studies_2004}. As a result, specimen preparation and pre-treatment practices (e.g., grinding, polishing, cleaning, and coating deposition) are expected to critically affect experimental results, yet these procedures vary significantly between laboratories, and there is currently no consensus on the best practices for ensuring a reproducible and oxide-free surface. The influence of gas impurities, such as oxygen contamination, can also be profound. Experiments have shown that small oxygen concentrations ($\sim$1000 vppm) can suppress the effect of hydrogen in fatigue crack growth \cite{somerday2013elucidating,komoda2014effect}. This has triggered significant interest in establishing hydrogen embrittlement test protocols that minimise contamination and therefore deliver conservative estimates of ductility, fracture toughness, and fatigue crack growth rates \cite{nibur2024non}. In the context of gaseous permeation studies, this highlights the importance of minimising the residual gas background in the cell (particularly, water vapour and oxygen), an aspect that has received limited attention. Nevertheless, impurities are likely to be present in most hydrogen storage and transport applications and, therefore, there is a need to quantitatively resolve the role that they play in hydrogen uptake. The use of modern MS-based detection systems enables simultaneous monitoring of multiple gas species and provides a powerful tool to quantify such effects.\\

Based on the above, this work aims to establish a robust gaseous hydrogen permeation methodology and systematically evaluate how surface condition and experimental variables (oxygen, temperature, pressure, etc.) influence the measured hydrogen permeation response. To this end, a dedicated permeation system combining high-sensitivity mass spectrometry (MS) detection with highly controlled sample preparation and environmental conditions is developed. The downstream MS detection side enables measurements of very low hydrogen fluxes with high temporal resolution, enabling accurate characterisation of transient permeation behaviour and extraction of transport parameters over a wide range of pressures and temperatures. Particular emphasis is placed on identifying and quantifying the impact of testing variables - including surface state and residual gas background - on hydrogen permeation measurements. Annealed pure Fe is employed as a model material for all experiments, enabling a more direct interpretation of hydrogen uptake and transport behaviour by reducing the influence of trapping effects and microstructural complexity as well as the comparison of results to a robust literature database \cite{volklDiffusionHydrogenMetals1978,kiuchi_solubility_1983}.  The work presented here provides a methodological framework to minimise experimental artefacts and improve the reproducibility of gaseous hydrogen permeation measurements, establishing a foundation for future systematic investigations of more complex phenomena, including the role of surface oxides, their interaction with mechanical loading, and the influence of gas impurities (e.g., oxygen) on hydrogen uptake.


\section{A high-sensitivity gas permeation system}
\label{Sec:Experimental}

We proceed to describe the characteristics of the high-sensitivity H$_2$ permeation system developed, beginning with a general background of hydrogen permeation (Section \ref{Sec:Basics}), then proceeding to describe the specifics of the assembled system (Section \ref{Subsec:GasPermeationApparatus}), followed by addressing calibration and detection sensitivity (Section \ref{Subsec:Calibration}), and finally detailing the testing protocol itself (Section \ref{Subsec:TestingProtocol}).

\subsection{Background: the basics of hydrogen permeation}
\label{Sec:Basics}

In H$_2$ permeation setups, a thin membrane of material (i.e., the specimen) separates two chambers: an upstream side exposed to hydrogen gas at a known pressure, and a downstream side where hydrogen emerging from the material is detected - see Fig. \ref{Fig2_PermeSystem}. The goal is to measure how readily hydrogen molecules dissociate, diffuse through the material as atoms, and recombine on the opposite surface. The recorded hydrogen flux $J$ as a function of time typically shows an initial transient regime that eventually reaches a steady-state permeation rate. Analysis of this behaviour allows determination of key transport parameters, including the diffusivity, solubility, and permeability of hydrogen in the material.\\

\begin{figure}[ht!]
    \centering
    \includegraphics[width=0.88\linewidth]{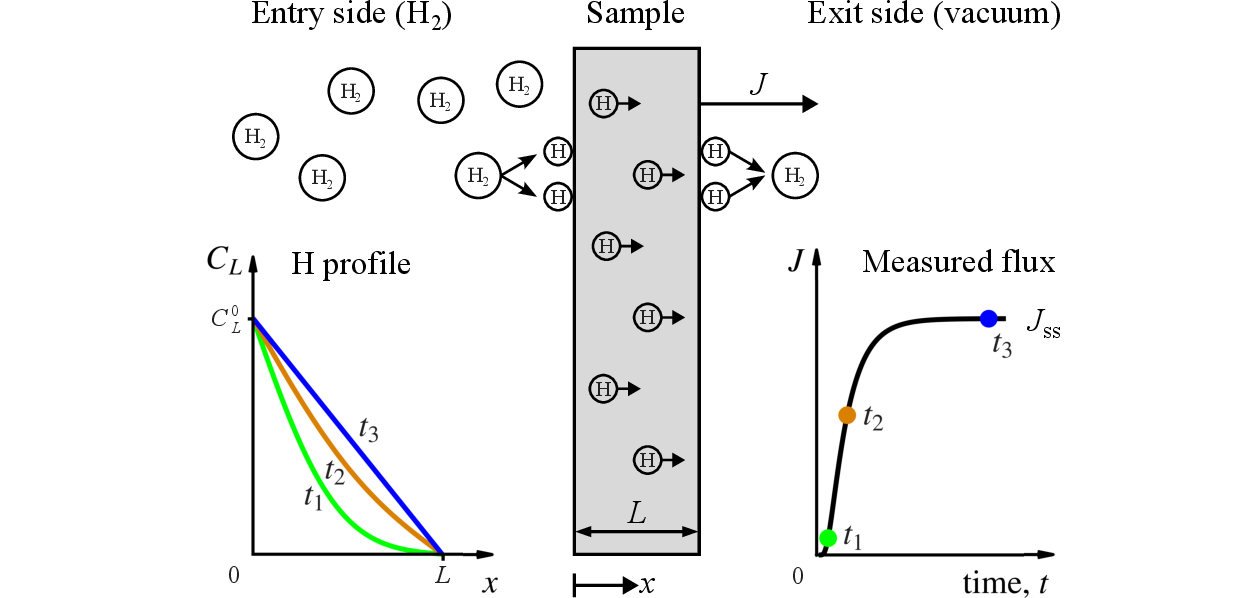}
    \caption{Schematic illustration of hydrogen permeation through a metal. Molecular hydrogen (H$_2$) at the entry side dissociates into atomic hydrogen, which is absorbed into the metal and diffuses across the sample of thickness $L$ driven by the concentration gradient. At the exit side (vacuum), hydrogen atoms recombine to form H$_2$ and desorb, giving rise to a measurable permeation flux $J$. The lower-left plot shows the evolution of the lattice hydrogen concentration $C_L(x,t)$ within the sample at different times ($t_1$, $t_2$, $t_3$). The lower-right plot shows the corresponding transient permeation flux $J(t)$, which increases with time and eventually approaches a steady-state ($J_{ss}$).}
    \label{Fig1_BriefBackground}
\end{figure}

The problem is effectively one-dimensional, with the hydrogen concentration being fixed by the hydrogen pressure on the entry side (also known as the high-pressure side, HPS) and maintained close to zero on the exit side (low-pressure side, LPS), generating a unidirectional hydrogen flux. Hydrogen atoms move from the entry side to the exit side \textit{via} interstitial diffusion, with defects within the sample, such as dislocations and grain boundaries, acting as sinks that \emph{trap} hydrogen atoms and thus slow down diffusion \cite{cupertino2023hydrogen}. As such, the total hydrogen concentration $C$ can be additively decomposed into a lattice hydrogen concentration $C_L$ and a trapped hydrogen concentration $C_T$. Since hydrogen transport is assumed to occur purely through interstitial lattice sites (i.e., traps are not interconnected), the hydrogen flux is defined as,
\begin{equation}\label{eq:Flux}
J = - D_L \frac{\partial C_L}{\partial x}
\end{equation}

\noindent with $D_L$ being the lattice hydrogen diffusion coefficient and $x$ the spatial coordinate. For the case of pure H$_2$ and relatively low pressures, the lattice hydrogen concentration at the entry side ($C_L^0$) can be related to the H$_2$ pressure ($p_{H_2}$) through Sieverts' law:
\begin{equation}\label{eq:Sievert}
    C_L^0 = S_L \sqrt{f_{H_2}}  \approx S_L \sqrt{p_{H_2}} 
\end{equation}

\noindent where $f_{H_2}$ is the hydrogen fugacity and $S_L$ is the hydrogen lattice solubility. For gas mixtures $C_L^0 = S_L \sqrt{p_{H_2}} = S_L \sqrt{x_{H_2} p}$, with $x_{H_2}$ being the mole fraction of hydrogen gas in the gas mixture and $p$ the total pressure of the mixture. For pressures high enough to sufficiently deviate from the ideal gas behaviour ($p_{H_2} > 100$ bar), the H$_2$ fugacity is no longer approximately equal to the H$_2$ pressure, and thus $C_L^0$ is estimated from $f_{H_2}$, using its relationship to the H$_2$ pressure, as per the Abel-Noble equation \cite{san2007permeability}. \\

The hydrogen concentration in the sample evolves with time, as shown in Fig. \ref{Fig1_BriefBackground}, until the steady-state condition is reached. At steady-state, the concentration profile across the membrane is described by a constant gradient; accordingly, considering Eq. (\ref{eq:Flux}), the steady-state flux reads,
\begin{equation}\label{eq:Jss}
    J_{ss} = \frac{D_L C_L^0}{L} 
\end{equation}

\noindent where $L$ is the membrane thickness. Eq. (\ref{eq:Jss}) emphasises the fact that trapping does not influence the magnitude of the steady-state flux (only the time at which it is achieved). The permeability is defined as the product of the diffusion coefficient and the solubility; accordingly, considering Eqs. (\ref{eq:Sievert}) and (\ref{eq:Jss}), the permeability coefficient can be expressed as
\begin{equation}\label{eq:Perme}
    P=D_L S_L = \frac{J_{ss} L}{\sqrt{p_{H_2}}}
\end{equation}

In practice, due to the role of trapping, it is difficult to experimentally measure lattice quantities. However, as elaborated in \ref{Appendix:Apparent}, under conditions of low trap occupancy, the concentration of trapped hydrogen is proportional to $C_L$ and therefore the following equivalent relationships can be established,
\begin{equation}\label{eq:AppEqs}
   J = -  D_{\text{app}} \frac{\partial C}{\partial x} \,\, , \,\,\,\,\,\,\,\,\,\, C^0 = S_{\text{app}} \sqrt{p_{H_2}} \,\, , \,\,\,\,\,\,\,\,\,\, J_{ss}=\frac{D_{\text{app}}C^{0}}{L}
\end{equation}

\noindent in terms of the total hydrogen concentration $C=C_L+C_T$, and apparent diffusivity ($D_{\text{app}}$) and solubility ($S_{\text{app}}$) variables, which account for both trapping and lattice contributions. For the case of the permeability coefficient, 
its magnitude does not distinguish between lattice and apparent conditions since $P_{\text{app}} = D_{\text{app}} S_{\text{app}}=D_L S_L = P$ (see \ref{Appendix:Apparent}). It is important to emphasise that the relationships in Eq. (\ref{eq:AppEqs}) only hold for the particular case where the trap occupancy is very small ($\theta_T < < 1$) - this assumption can be verified by assessing whether the permeation transient conforms to Fick's law \cite{oriani1970diffusion,turnbull2015perspectives}; if the permeation transient is steeper than projected by Fick's law, then numerical analysis using a diffusion-trapping model is needed \cite{diaz2025comsol}. If Fick's law provides a good fit to the data, a common protocol is to obtain the apparent diffusion coefficient from the transient flux response. For example, using the so-called time-lag method:
\begin{equation}
  D_{\text{app}}=\frac{L^2}{6t_\text{lag}} 
\end{equation}

\noindent where $t_\text{lag}$ corresponds with the time at which the measured flux equals $J(t)=0.63J_{ss}$. Or, as done in this work, by fitting Fick's law to the transient flux versus time response;
\begin{equation}\label{eq:JtFou}
J(t)=J_\mathrm{ss}\left[\,1+2\sum_{n=1}^{\infty}(-1)^n
\exp\!\left(-\frac{n^2\pi^2 D_\mathrm{app}}{L^2}\,t\right)\right]  
\end{equation}

Then, the measured steady-state flux can be used to determine the permeability, as per Eq. (\ref{eq:Perme}), and subsequently, the apparent solubility is calculated as $S_{\text{app}}=P/D_{\text{app}}$.

\subsection{System description}
\label{Subsec:GasPermeationApparatus}

\begin{figure}[H]
    \centering
    \includegraphics[width=1\linewidth]{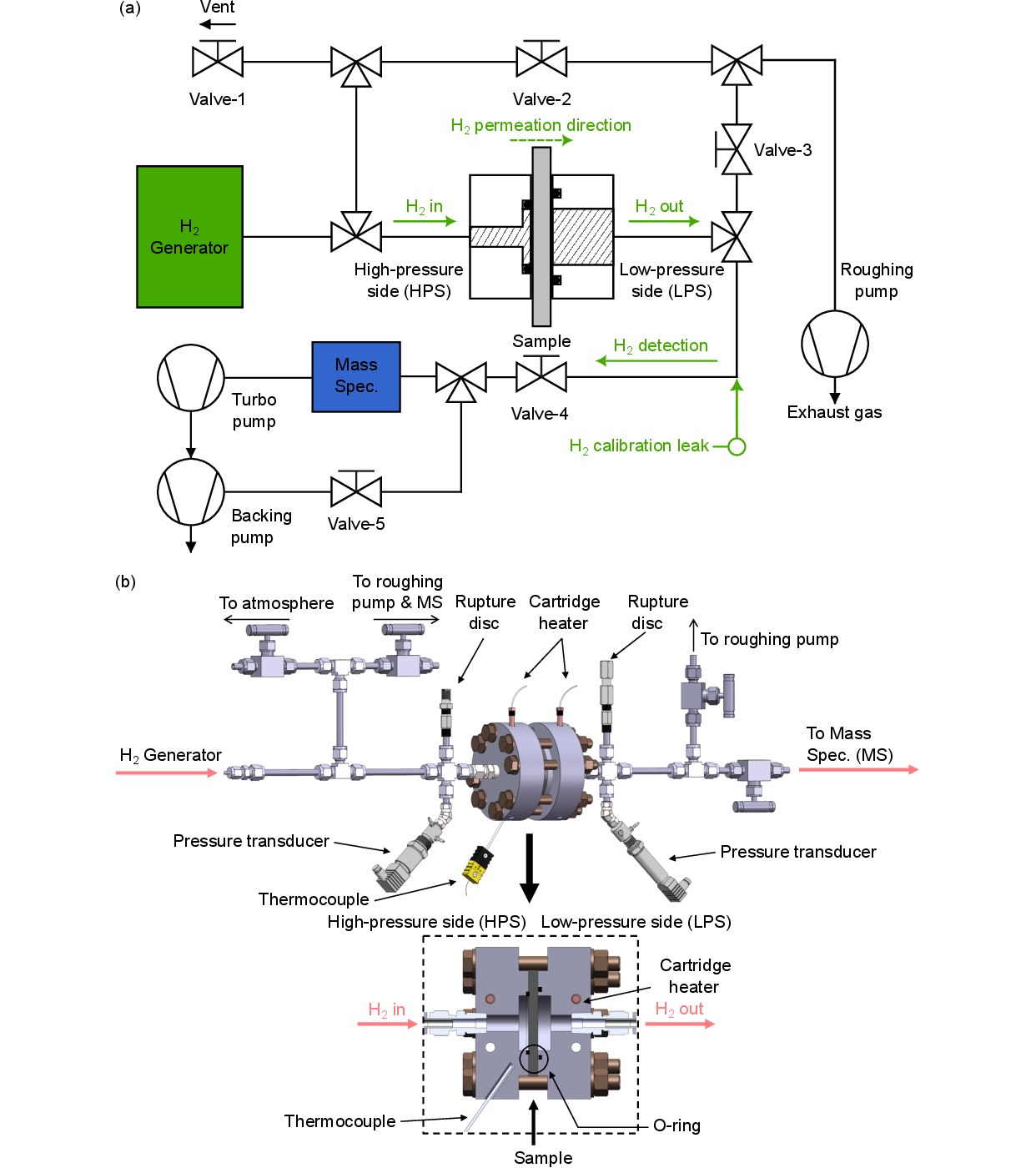}
    \caption{A high-resolution H$_2$ permeation system: (a) Schematic representation of the system, including the hydrogen generator, high-pressure side (HPS), low-pressure side (LPS), mass spectrometer (MS), roughing and turbomolecular pumps, and calibration leak line. Valves controlling gas admission, evacuation and isolation of the LPS during testing are indicated. (b) Three-dimensional layout of the permeation cell and its main components, including rupture discs, pressure transducers, cartridge heaters, thermocouples, and connections to the roughing pump and MS. The inset shows a cross-sectional view of the cell assembly, highlighting the sample position and O-ring sealing configuration.}
    \label{Fig2_PermeSystem}
\end{figure}

The high-sensitivity hydrogen gas permeation system developed in this work is comprised of: (i) a high-purity hydrogen source, (ii) a custom stainless steel permeation cell, (iii) an ultra-high vacuum (UHV) downstream detection line coupled to a quadrupole mass spectrometer (MS), and (iv) associated gas delivery and vacuum pumping systems. A schematic of the hydrogen permeation setup is shown in Fig. \ref{Fig2_PermeSystem}(a), and the three-dimensional layout of the permeation cell is presented in Fig. \ref{Fig2_PermeSystem}(b).\\

Hydrogen is supplied from on-site electrolysed water using an LNI SWISSGAS HG PRO 1100 hydrogen generator, capable of producing ultrapure hydrogen (99.99999\%) at pressures up to 16 bar. Alternatively, pure hydrogen or hydrogen mixtures containing purposeful impurities such as O$_2$, CO$_2$, and CH$_4$ can be supplied at pressures up to 50 bar directly from certified gas cylinders. Upstream pressure is monitored using a transducer rated for up to 69 bar. Note that, in this work, hydrogen pressure is expressed as gauge pressure unless otherwise explicitly stated. For example, a hydrogen pressure of 5 bar refers to 5 bar gauge pressure, which corresponds to approximately 6 bar absolute pressure (1 bar atmospheric pressure + 5 bar gauge pressure). Gas lines consist of 1/4$''$ and 1/8$''$ 316L stainless steel tubing, valve and fitting components, arranged compactly to minimise dead volume and ensure leak-tight operation. Gas charging and evacuation are controlled using Swagelok valves, enabling evacuation, venting, and accurate pressurisation.\\

The permeation cell, shown in Fig. \ref{Fig2_PermeSystem}(b), consists of a customised Parr A2000G design constructed from 316 stainless steel, rated for a pressure of 50 bar at 275 $^\circ$C. The cell is equipped with rupture discs on both sides for overpressure protection and is heated using four embedded 200 W cartridge heaters. Temperature is regulated by a Parr 4848B PID controller using a type-K thermocouple embedded within the assembly. The specimen is placed at the centre of the cell and sealed using M10 bolts. Sealing is achieved using MBMS perfluoroelastomer (FFKM) O-rings, selected for their chemical inertness and high-temperature resistance (up to 330 $^\circ$C). New O-rings are used for each experiment to ensure reproducible sealing. Viton O-rings were also tested and showed similar behaviour at low temperatures but signs of degradation at higher temperatures ($\textgreater$ 100 $^\circ$C). 

The entry side (HPS) volume is $\approx$20 cm$^3$, and the effective permeation area is $\approx$10 cm$^2$. The downstream LPS serves as the hydrogen detection volume and is maintained under UHV conditions ($\approx10^{-9}$ mbar at room temperature) throughout the experiment, using a turbomolecular pump. An auxiliary pressure transducer rated to 0.34 bar (with precision of 0.00085 bar) is installed on the LPS for redundancy and for potential operation in manometric mode when fluxes exceed the MS detection range (a scenario not considered in this study). In this work, hydrogen detection is performed using a HAL/3F RC PIC quadrupole MS (Hiden Analytical Ltd.), operating under the same UHV conditions. The instrument employs a triple-filter quadrupole analyser coupled with a pulse ion counting detector, providing a detection limit of $5 \times 10^{-16}$ mbar in terms of equivalent partial pressure.  The system exhibits a linear dynamic range exceeding seven orders of magnitude in ion count rate (counts/s), which is the quantity used for hydrogen quantification in this work. Fast data acquisition rates (up to 1000 Hz) enable accurate monitoring of rapid transient events. Finally, while not relevant for the low pressures considered in this study, a backing plate can be installed in the LPS to mitigate mechanical deformation of samples exposed to high pressures.

\subsection{Calibration and sensitivity}
\label{Subsec:Calibration}

In order to provide an accurate quantification of the hydrogen flux measured by the MS during a permeation experiment (which is detected as a counts/s raw signal), it is necessary to calibrate the instrument using a known hydrogen flux. Hence, calibration was conducted at room temperature by measuring a known flux of pure hydrogen gas (m/z = 2) provided by a certified calibration leak cylinder supplied by LACO Technologies. The quadrupole MS allows adjustment of the ion source emission current between 1 and 2000 \textmu A, which controls the electron flux available for ionising the gas species and thus determines the detection sensitivity. Based on previous experience, a 20 \textmu A emission current was selected as optimal in order to obtain a high count rate gain at small fluxes and thus high sensitivity, without risking exceeding the count threshold of the detector and introducing excessive noise.\\ 

\begin{figure}[H]
    \centering
    \includegraphics[width=1\linewidth]{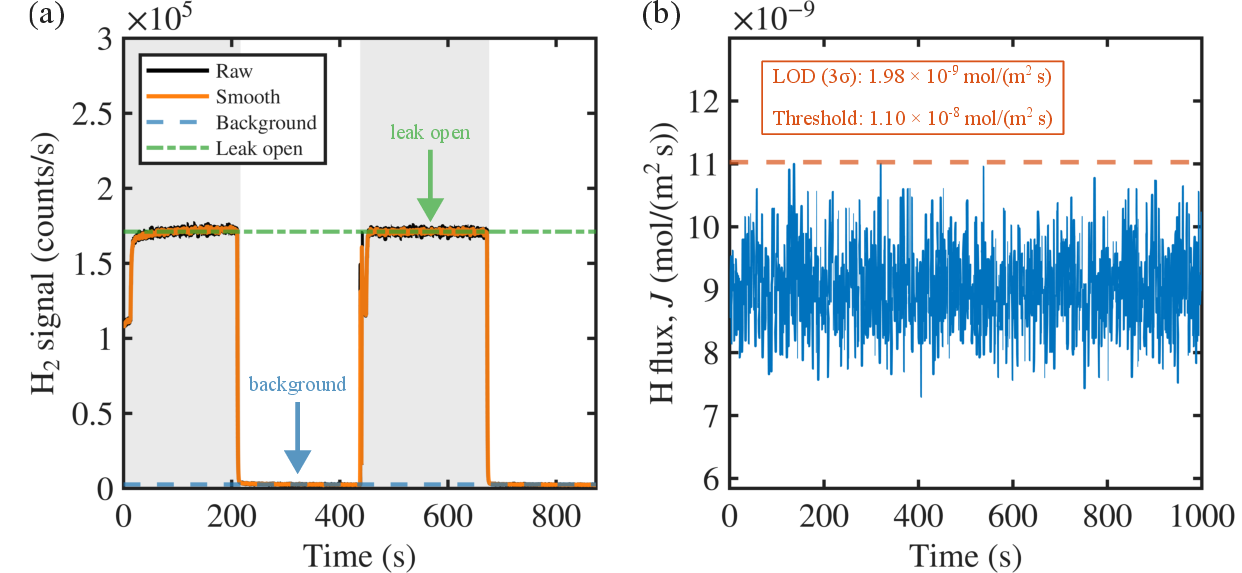}
    \caption{Calibration and sensitivity assessment of the quadrupole mass spectrometer (MS) used in the permeation system. (a) Calibration signal recorded at an ion source emission current of 20 $\mu$A. The raw signal (blue) and smoothed trace (orange) are shown, together with the background level and the response obtained upon opening the certified hydrogen calibration leak. (b) Background hydrogen flux measured at room temperature under UHV conditions prior to permeation testing. The statistical limit of detection (LOD = 3$\sigma$) and the practical detection threshold ($\tau$=$\mu$+3$\sigma$) are indicated. Data were acquired with the MS directly connected to the permeation cell and shared vacuum line.}
    \label{Fig3_SensitivityLOD}
\end{figure}

During calibration, conducted prior to the experimental program, the certified leak cylinder is connected to the MS, as illustrated in Fig. \ref{Fig2_PermeSystem}(a), and the isolating valve is opened and closed alternatively to obtain signal plateaus corresponding to the equipment background (closed valve) and the calibration leak (open valve), as shown in Fig. \ref{Fig3_SensitivityLOD}(a). The stable cycle segments were then used to determine the calibration factor ($K_{cal}$ = 4.64 $\times$ 10$^{-15}$ mol H/count) by establishing an equivalence between the molar flow rate of hydrogen atoms provided by the certified leak cylinder ($\dot{n}_{leak}$ = 7.82 $\times10^{-10}$ mol H/s) and the background-subtracted hydrogen signal measured by the MS at a 20 \textmu A emission current ($I^{cal}_H$ = 168758 counts/s). Once $K_{cal}$ is determined, representing the number of moles of hydrogen atoms corresponding to one count recorded by the MS, conversion of the raw signal intensity during a permeation test ($I_H$) into hydrogen flux, $J$ (mol/(m$^2$ s)), is straightforward:  
\begin{equation}
    J = \frac{I_{H}  K_{cal}}{A_{\text{sample}}}
    \label{eq:fluxC}
\end{equation}

\noindent where $A_{\text{sample}}$ is the cross-sectional area of the sample being permeated by hydrogen. Throughout this work, the hydrogen flux $J$ always corresponds to the atomic hydrogen (e.g., H) flux. \\ 

To quantify the detection sensitivity of the gas permeation system, the statistical properties of the background signal were analysed after the system reached the steady-state condition. Fig. \ref{Fig3_SensitivityLOD}(b) shows the hydrogen signal measured by the MS on the LPS side of the permeation cell during a 1000 s period prior to the start of the permeation experiment. At this stage, the system has been fully pumped down and stabilised, thus no hydrogen pressure has yet been applied on the HPS, and both compartments are at the base vacuum level ($\approx10^{-9}$ mbar at room temperature).\\ 

The limit of detection (LOD) and the raw-signal detection threshold ($\tau$) were determined from the background noise measurements according to:
\begin{equation}
\mathrm{LOD}=3\sigma,\quad \tau=\mu+3\sigma
\end{equation}

\noindent where $\mu$ and $\sigma$ are the mean value and standard deviation of the background signal, respectively. The LOD characterises the noise-limited sensitivity of the system, quantifying the smallest signal that can be distinguished from random background fluctuations, independent of any baseline offset. In contrast, $\tau$ defines a practical detection threshold, corresponding to the signal level that must be exceeded in the raw data to be confidently interpreted as a real hydrogen permeation signal above the background. From these background statistics, a flux LOD of 3$\sigma$ = 1.98 $\times$ 10$^{-9}$ mol H/(m$^2$ s) was obtained, together with a corresponding detection threshold of $\tau$ = 1.10 $\times$ 10$^{-8}$ mol H/(m$^2$ s). It should be noted that the signal-to-noise ratio depends strongly on the magnitude of the permeation flux. For experiments conducted at room temperature and low hydrogen pressures, such as the 1–5 bar tests reported here, the measured fluxes can approach the detection limit of the system, leading to comparatively larger relative noise. In contrast, when the flux is increased, for example by testing at a higher temperature or pressure, the signal-to-noise ratio improves substantially. The methodology can also be adapted to improve the signal-to-noise ratio by reducing the sample thickness, increasing the exposed area, extending the acquisition time, and applying suitable MS signal averaging.\\

\subsection{Testing protocol and data analysis}
\label{Subsec:TestingProtocol}

\begin{figure}[H]
    \centering
    \includegraphics[width=0.95\linewidth]{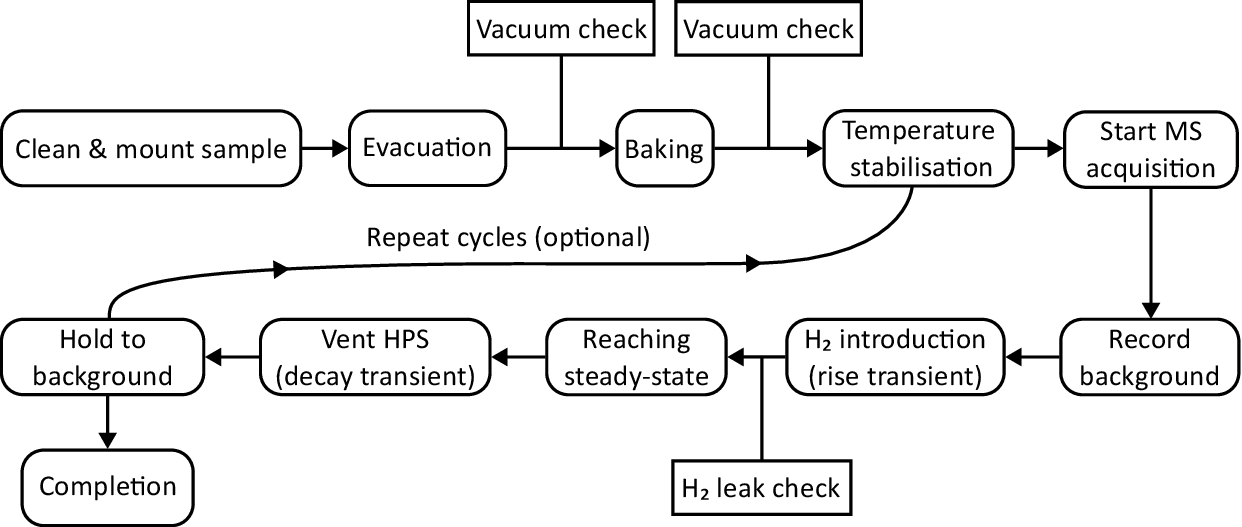}
    \caption{Flow chart of the measurement protocol used for gas permeation testing. After cleaning and mounting the specimen, the system undergoes evacuation, bake-out and vacuum verification prior to temperature stabilisation and background acquisition. Permeation is initiated by hydrogen pressurisation of the HPS (rise transient), followed by steady-state monitoring and subsequent venting and evacuation (decay transient). The rise–decay sequence may be repeated under different testing conditions before final completion of the experiment.}
    \label{Fig4_ProcessFlow}
\end{figure}

Due to the lack of standardisation in gas permeation testing, a measurement protocol was established based on previous literature \cite{young_measurement_2020,chalfoun_hydrogen_2022,bryan_diffusivity_1963,koren_experimental_2023,mendibide_effect_2024} and on extensive preliminary trials, to ensure maximum repeatability and reliable baseline conditions. Each experiment follows the procedure illustrated in Fig. \ref{Fig4_ProcessFlow}. \\

After sample preparation and cleaning - a critical factor influencing gas permeation results, as discussed in detail in Section \ref{Sec:MethResults} - the specimen is mounted in the permeation cell using fresh O-rings for each test (except for room-temperature tests, where O-rings may be reused). Both the LPS and HPS compartments are then evacuated using the roughing pump to $\approx10^{-7}$ mbar. An initial vacuum integrity check is performed by monitoring pressure stability before bake-out. The assembly is subsequently baked at 150 $^\circ$C for approximately 15 h to minimise residual hydrogen and other desorbable species (e.g., water and oxygen) released from the cell, O-rings or sample \cite{miller_permeation_1975, chalfoun_hydrogen_2022, rothig_gaseous_2025}. This temperature corresponds to the maximum testing temperature used in this study, ensuring that the lowest attainable baseline for residual gases is achieved. After bake-out, the system is cooled under continuous pumping. Once room temperature is reached, both compartments are opened to the turbomolecular pump of the MS system, allowing high-vacuum conditions of approximately $10^{-9}$ mbar. A second vacuum integrity check is then conducted under UHV conditions.\\ 

The cell is subsequently brought to the desired testing temperature (or maintained at room temperature), and the pressure is allowed to stabilise. Residual gas levels, including oxygen, water and nitrogen, are measured in both compartments to verify acceptable background conditions. The LPS is then isolated from the rest of the equipment (valve 3 in Fig. \ref{Fig2_PermeSystem}(a)), MS acquisition is initiated, and a background signal is recorded for 1-2 hours. Once a stable minimum H$_2$ background is achieved in the LPS, the permeation test is initiated by pressurising the upstream side with H$_2$, marking the beginning of the rise transient. Potential H$_2$ leaks are monitored at this stage using a gas sniffer (Panther Pro, ION Science Ltd., resolution of 1 ppm H$_2$). After steady-state flux is reached, the H$_2$ supply is interrupted, and the HPS chamber is evacuated through the roughing pump, initiating the decay transient. The rise-decay sequence may be repeated as required.

Prior to analysis, permeation data are processed to improve the signal-to-noise ratio while preserving transient features. Baseline correction, smoothing and downsampling are performed using a Savitzky-Golay algorithm implemented \textit{via} Matlab. To ensure convergence, $n=40$ Fourier terms are considered when fitting the flux transient, as per Eq. (\ref{eq:JtFou}). Parameter optimisation was performed using a gradient-based multi-start nonlinear least-squares procedure (Levenberg-Marquardt algorithm, Matlab). The optimal value of $D_{\text{app}}$ was obtained by minimising the residual error, and goodness of fit was assessed using the coefficient of determination (R$^2$).

\section{Impact of surface condition and testing variables on hydrogen permeation in pure iron}
\label{Sec:MethResults}

\subsection{Methods}
\label{Subsec:Methods}

\subsubsection{Material and sample dimensions}

All experiments were conducted on ARMCO\textsuperscript{\textregistered} iron samples with a reported purity of 99.98 wt.\% Fe, supplied by Goodfellow Ltd. as 1 mm-thick sheets in the annealed condition. This material was selected due to the extensive literature available on hydrogen gas permeation in pure Fe \cite{volklDiffusionHydrogenMetals1978, kiuchi_solubility_1983}, as well as its intrinsically low trapping density. Additionally, the single-phase ferritic structure, uniform grain distribution, and minimal impurity content reduce the influence of microstructural trapping, thereby enabling isolation of surface effects and testing variables. For the gas permeation experiments, 70 mm diameter discs were extracted from the sheet using electrical discharge machining (EDM). The initial sample thickness, prior to grinding, was 1 mm.

\subsubsection{Sample surface preparation}

There is no consensus in the community on the optimal surface preparation for gas permeation samples, which is needed to ensure reliable and reproducible results with minimal influence of surface oxides or other artefacts. Therefore, surface preparation was systematically varied herein on both the HPS and LPS of the Armco iron samples; Table \ref{tab:Surface treatment matrix} summarises the surface preparation combinations investigated. Surface preparation on the HPS included mechanical grinding (Fe), acid pickling (p), and electrodeposition of a nominally 100 nm-thick Pd layer (Pd). For example, a sample with ID Pd/p/Fe/p/Pd corresponds to a sample with Pd coatings on both the HPS and LPS, with the Fe substrate previously ground and acid pickled. Pd is an excellent catalyst, minimising surface effects, while acid pickling is done to remove any oxide layers present. In our experimental campaign, we chose to Pd coat the LPS, where hydrogen detection takes place, whereas the raw Fe surface on the HPS was, in some cases, directly exposed to H$_2$. However, an additional experiment, reported in \ref{Appendix:B}, was conducted to investigate the condition in which a Pd layer is absent in the LPS. For the results reported in the main text, gas permeation experiments were conducted at 25 $^\circ$C and 5 bar H$_2$ for the four surface conditions shown in Table \ref{tab:Surface treatment matrix}. If no permeation signal was detected within the first hour, measurement times were extended to between 16 and 22 h. Duplicate tests were performed to assess repeatability. 

\begin{table}[H]
\centering
\caption{Surface preparation matrix used to investigate the influence of surface condition on hydrogen permeation. “Fe” denotes a ground iron surface, “p” indicates acid pickling in HCl, and “Pd” corresponds to electrodeposition of an approximately 100 nm-thick Pd layer. Sample IDs describe the sequence of surface treatments applied from the high-pressure side (HPS) to the low-pressure side (LPS).}
\begin{tabular}{c c c c}
\hline
Sample ID & High Pressure Side & Low Pressure Side \\
\hline
Fe/Pd & Fe & Pd \\
Fe/p/Pd & Fe & pickling + Pd \\
Pd/Fe/Pd & Pd & Pd \\
Pd/p/Fe/p/Pd & pickling + Pd & pickling + Pd \\
\hline
\end{tabular}
\label{tab:Surface treatment matrix}
\end{table}

Grinding was performed on both surfaces of all samples as a baseline treatment, using successive SiC papers up to 2500 grit, resulting in a final thickness of approximately 0.7 mm. Because freshly exposed Fe oxidises rapidly in air \cite{roosendaal_passivation_1999,gilroy_oxidation_1965}, with native oxide formation initiating within very short ambient exposures and potentially reaching thicknesses of the order of 1--5 nm over minutes to hours depending on the environment \cite{krugerRoomTemperatureOxidation1964,grosvenorUseQUASESXPS2004,bhargava_characterization_2007}, a strict protocol was followed to remove residues and limit reoxidation. This involves ultrasonic cleaning in acetone for 5 min, rinsing in deionised (DI) water for 2--3 seconds, and immediate drying with N$_2$. For Pd coating, samples were transferred immediately into the electroplating solution after grinding, with the time between surface preparation and immersion in the plating bath kept below 5 s. Pd was deposited from a commercial plating solution (Palladium Bath JE42D, Jentner, 2 g/L Pd) at a constant current density of $\sim$2 mA/cm$^2$ for 15-20 min, yielding a thickness of $\sim$100 nm \cite{zafra_comparison_2022}. After coating, samples were rinsed in DI water, dried with N$_2$, and stored in a desiccator until testing. In some cases, acid pickling in HCl was applied between grinding and Pd coating to further remove surface oxides \cite{chalfoun_hydrogen_2022, mendibide_effect_2024} via immersion in an 18\% HCl solution at room temperature for 10-–20 s, followed by immediately rinsing with DI water for 5--10 seconds to remove any residual acid, and direct transfer to the Pd electroplating solution within less than 5 s. Control experiments confirmed that, provided exposure to air between preparation steps was minimised, surface preparation conducted in air yielded permeation results consistent with those obtained under inert atmosphere conditions, i.e., preparation inside a nitrogen-filled glovebox, indicating that the adopted protocol was effective in preventing measurable reoxidation before Pd deposition/testing.

\subsubsection{Temperature and hydrogen reduction treatment}

In addition to the tests outlined above, the influence of hydrogen reduction treatment on permeation behaviour was investigated via experiments conducted on samples with different surface conditions (Fe/Pd, Fe/p/Pd, Pd/Fe/Pd, and Pd/p/Fe/p/Pd). The temperature was sequentially increased from 25 $^\circ$C to 150 $^\circ$C in 10-25 $^\circ$C increments while maintaining a constant hydrogen pressure of 5 bar on the HPS. At each temperature, the background signal in the LPS was allowed to stabilise before introducing H$_2$ at 5 bar, which was maintained for at least 30 minutes. If no signal exceeding the defined LOD was observed within this period, hydrogen permeation was considered undetectable at that temperature. After each step, hydrogen was vented, and both sides of the cell were evacuated before increasing the temperature. This procedure was repeated until a measurable permeation signal was obtained, enabling identification of the temperature at which hydrogen permeation becomes detectable due to progressive reduction of surface oxides formed during specimen preparation. After reaching 150 $^\circ$C, three consecutive rise-decay transients were performed at 5 bar. This approach enabled direct evaluation of the evolution of $D_{\text{app}}$ across successive transients, thereby capturing the progressive changes in hydrogen transport associated with oxide reduction. At this temperature, trapping effects are expected to be negligible  \cite{chen_hydrogen_2025}; therefore, variations in $D_{\text{app}}$ between consecutive transients can be primarily attributed to surface activation resulting from the gradual reduction of oxide layers. This protocol differs from conventional approaches, where oxide reduction is typically performed at constant temperature for a fixed duration prior to permeation testing \cite{rothig_gaseous_2025}. In contrast, the present method allows in-situ monitoring of the evolution of transport parameters during the reduction process, providing additional insight into the kinetics of surface activation. Subsequently, the system was cooled stepwise in 25 $^\circ$C intervals until room temperature was reached. At each temperature, two rise-decay cycles were conducted. This procedure enabled assessment of whether hydrogen-reduced surface conditions were retained upon cooling and whether measurable permeation persisted at room temperature.

\subsubsection{Residual gas background}

To investigate the influence of residual gas background on hydrogen permeation, experiments were performed at 25 $^\circ$C and 5 bar H$_2$ using different sealing configurations on 0.7 mm-thick samples of Armco iron with identical surface condition (Pd/p/Fe/p/Pd), thereby minimising the influence of surface oxide variations. Three configurations were investigated: new seals under ambient conditions, conditioned seals under ambient conditions, and conditioned seals in a N$_2$-purged environment. Conditioned O-rings correspond to seals previously exposed to high-temperature experiments (150 $^\circ$C), while the inert purge condition was achieved by placing the cell inside a glovebox continuously flushed with high-purity N$_2$. These configurations were selected to intentionally vary the level of residual gas contamination (O$_2$, H$_2$O, and N$_2$) before hydrogen exposure. Each experiment consisted of four consecutive rise-decay permeation transients conducted under identical conditions. The residual gas background in the HPS was characterised using MS immediately before hydrogen admission for each transient. The reported background corresponds to the average signal of H$_2$O (m/z = 18), N$_2$ (m/z = 28), and O$_2$ (m/z = 32) measured over the 1–2 h period preceding each transient. As the MS was not calibrated for these species, values are reported in counts/s. No oxygen or other gases were intentionally added to the hydrogen stream; all experiments were conducted using high-purity (99.99999\%) H$_2$ supplied by an electrolyser. This approach enables systematic evaluation of how sealing conditions and environmental control during evacuation influence residual gas background and, consequently, the measured permeation response.

\subsubsection{Varying pressure and temperature}

To assess the influence of H$_2$ pressure on the permeation response ($J_{ss}$ and $D_{\text{app}}$) and verify bulk-controlled diffusion (i.e., $J_{ss}$ following Sieverts’ law), permeation transients were obtained at 25 $^\circ$C over a pressure range from 1 to 5 bar using samples with the optimised Pd/p/Fe/p/Pd surface condition, where surfaces effects are minimised. A single permeation experiment was conducted in which consecutive rise-decay cycles were performed at 1 bar (two transients), followed by 2, 3, 4, and 5 bar. Subsequent transients at the same H$_2$ pressure are often used to reduce the influence of trapping through non-equilibrium phenomena such as slow detrapping (`irreversible trapping') \cite{Nelson1973}. In addition to pressure, temperature is a key variable governing hydrogen transport in metals, as it influences both hydrogen mobility within the lattice and its equilibrium solubility. Permeation experiments were therefore conducted at a pressure of 5 bar and temperatures of 25, 50, 100 and 150 $^\circ$C using the same optimised surface condition (Pd/p/Fe/p/Pd). At each temperature, several consecutive rise-decay transients were recorded, after which the sample was cooled, removed from the cell, cleaned, and reused for subsequent experiments at higher temperatures. Analysis of the temperature dependence of the permeation response enables the determination of activation energies associated with hydrogen transport (via Arrhenius relationships) and facilitates comparison with literature values.

\subsection{Results}
\subsubsection{Influence of surface preparation}

\begin{figure}[H]
    \centering
    \includegraphics[width=0.9\linewidth]{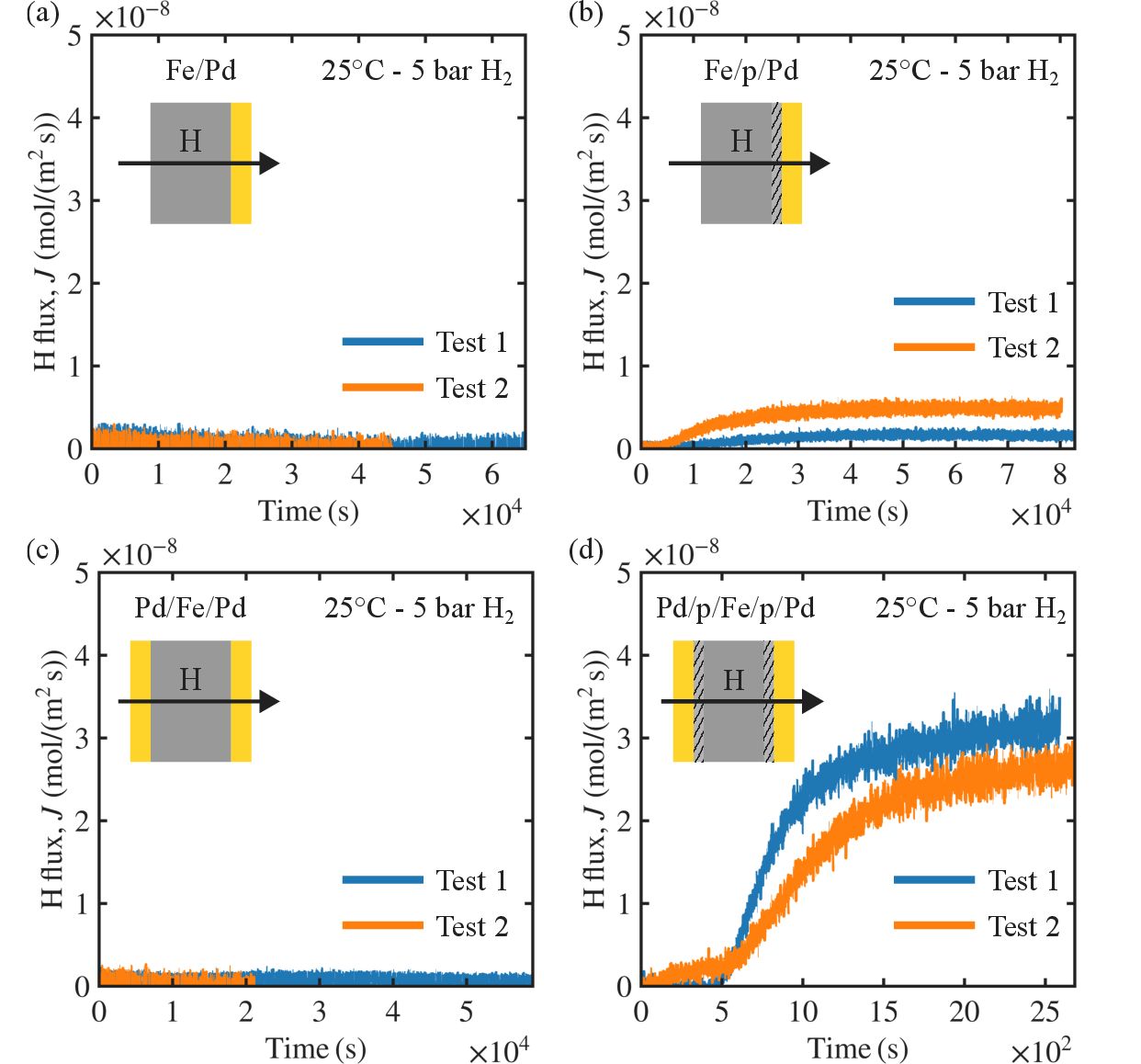}
    \caption{Hydrogen permeation transients measured at 25 $^\circ$C and 5 bar for different sample surface preparations: (a) Fe/Pd, (b) Fe/p/Pd, (c) Pd/Fe/Pd, and (d) Pd/p/Fe/p/Pd. No detectable permeation is observed for Fe/Pd and Pd/Fe/Pd (oxide present on both sides), whereas Fe/p/Pd exhibits a delayed, low-intensity response (single oxide layer) and Pd/p/Fe/p/Pd shows a clear transient and steady-state flux (oxide removed on both sides).}
    \label{Fig5_PdCoatingComparision}
\end{figure}

The hydrogen permeation transients measured at 25 $^\circ$C and 5 bar for four different surface treatments (Fe/Pd, Fe/p/Pd, Pd/Fe/Pd, and Pd/p/Fe/p/Pd) are shown in Fig. \ref{Fig5_PdCoatingComparision}. A clear distinction in permeation response is observed, with three regimes depending on surface preparation. For the Fe/Pd (Fig. \ref{Fig5_PdCoatingComparision}(a)) and Pd/Fe/Pd (Fig. \ref{Fig5_PdCoatingComparision}(c)) configurations, the measured signal remained well below the instrumental LOD (1.98$\times$10$^{-9}$ mol/(m$^{2}$s)) throughout the experiment, even for durations up to 18 h, indicating negligible hydrogen permeation under these conditions. 

In contrast, the Pd/p/Fe/p/Pd condition (Fig. \ref{Fig5_PdCoatingComparision}(d)) exhibits a well-defined permeation transient, initiating after approximately 500 s of hydrogen exposure in the HPS and reaching a steady-state flux of approximately 3$\times$10$^{-8}$ mol/(m$^{2}$s) within less than 1 h, i.e., more than an order of magnitude above the LOD. This behaviour was reproducible across duplicate experiments and enabled determination of an apparent hydrogen diffusivity of $\sim$7.4$\times$10$^{-11}$ m$^{2}$/s. Comparatively, the Fe/p/Pd configuration (Fig. \ref{Fig5_PdCoatingComparision}(b)) shows intermediate behaviour, characterised by a delayed and low permeation response. In this case, a measurable signal is only observed after prolonged exposure ($\approx$8 h), with the steady-state flux remaining close to the detection limit. The corresponding average values ($D_{\text{app}}=5.46\times10^{-12}$ m$^2$/s and $J_{ss}=2.84\times10^{-9}$ mol/(m$^{2}$s), Table \ref{tab:PermeParameters}) are significantly lower than those obtained for the fully conditioned Pd/p/Fe/p/Pd condition. Similar behaviour has been reported by Mendibide et al. \cite{mendibide_effect_2024} for X65 pipeline steel with comparable surface preparation (polished, pickled, and Pd-coated only on the detection side), where very low permeation signals and noticeable scatter between replicates were observed under both dry and wet H$_2$ conditions. The observed variability was attributed to differences in the thickness and/or integrity of the surface oxide layer, which predominantly affect hydrogen entry and steady-state flux.\\

These results demonstrate that hydrogen permeation at room temperature is strongly governed by the surface condition. When only one surface is covered by an oxide layer, hydrogen can still permeate, although the process is significantly hindered and controlled by surface reaction kinetics. This behaviour is further supported by the additional experiment presented in \ref{Appendix:B}, in which only the entry surface was pickled and Pd-coated while the detection side remained oxidised. In this case, hydrogen permeation is observed due to the catalytic activity of Pd and the apparently limited resistance of the oxide to bulk diffusion; however, the signal does not reach steady state even after more than 25 h, indicating that hydrogen recombination at the oxidised exit surface is severely hindered. In contrast, when oxide layers are present on both sides of the Fe membrane (Fe/Pd and Pd/Fe/Pd), hydrogen permeation is effectively suppressed, demonstrating that the combined effect of two oxide interfaces constitutes a strong kinetic barrier to hydrogen transport. Consequently, reliable and measurable hydrogen permeation requires both oxide removal and Pd coating on the entry and exit surfaces. While the Pd layer limits further oxidation and promotes hydrogen dissociation at the entry surface and recombination at the exit surface \cite{girottoEffectPhysicochemicalProperties2023,manolatos_necessity_1995}, its effectiveness depends critically on the underlying surface state of the Fe membrane. In practice, even brief exposure to ambient conditions leads to the rapid formation of a native oxide layer. Although only a few nanometres thick, this oxide is persistent and can dominate hydrogen uptake by affecting adsorption, dissociation, and recombination processes \cite{grosvenor_studies_2004,roosendaal_passivation_1999,wampler_surface-limited_1986,wampler_hydrogen_1989}, thereby distorting the permeation response from ideal Fickian behaviour. The magnitude of this effect depends strongly on oxide chemistry and defect structure, with features such as porosity, cracking, and variations in Fe\(^{2+}\)/Fe\(^{3+}\) ratio introducing additional transport pathways or trapping sites \cite{houben_comparison_2019,vucko_hydrogen_2022}.\\  

\begin{figure}[ht!]
    \centering
    \includegraphics[width=1\linewidth]{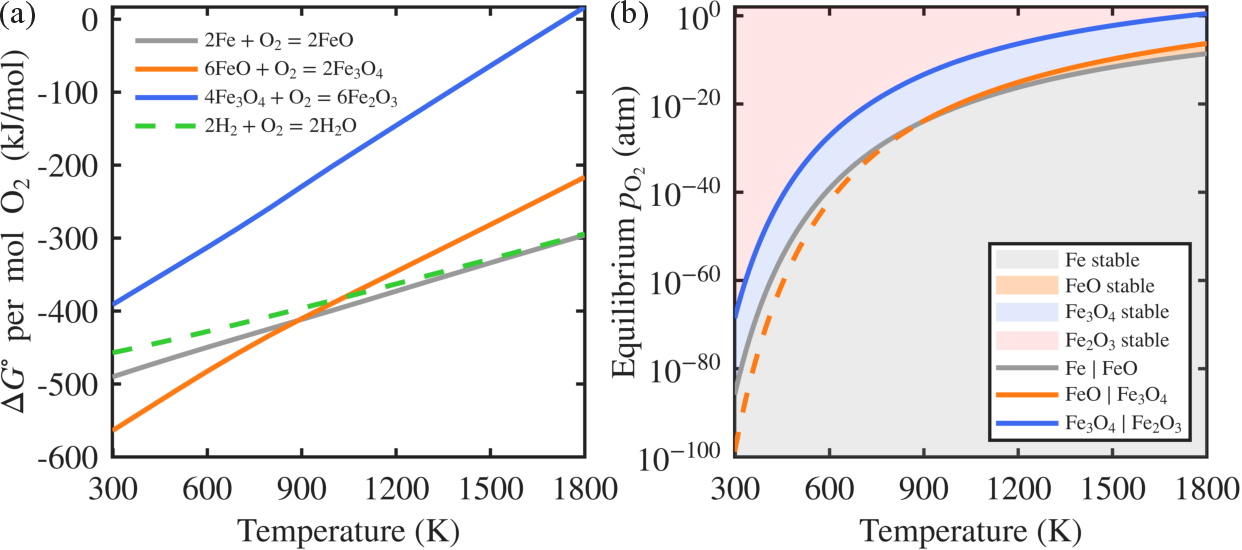}
    \caption{Thermodynamics of oxide formation in the Fe-O system. (a) Ellingham diagram showing the standard Gibbs free energy change ($\Delta G^\circ$) for Fe–O and H–O reactions as a function of temperature. (b) Equilibrium oxygen partial pressure ($p_{O_2}$) diagram indicating the stability domains of Fe, FeO, Fe$_3$O$_4$, and Fe$_2$O$_3$. Shaded regions denote thermodynamically stable phases. The dashed line represents the temperature range over which FeO is thermodynamically unstable. Thermodynamic data are taken from the NIST database \cite{linstromNISTChemistryWebBook1997,thomas_c_allison_nist-janaf_2013}. FeO corresponds to w\"ustite, Fe$_3$O$_4$ to magnetite, and Fe$_2$O$_3$ to hematite.}
    \label{Fig6_EllinghamDiagram}
\end{figure}  

From a thermodynamic perspective, iron oxidation is strongly favoured under ambient conditions. As shown in the Ellingham diagram (Fig. \ref{Fig6_EllinghamDiagram}(a)), the position of the H-O equilibrium line (dashed line) above two Fe-O lines at room temperature ($\sim$300 K), indicates that oxide formation is thermodynamically preferred over hydrogen reduction. Furthermore, the phase stability diagram (Fig. \ref{Fig6_EllinghamDiagram}(b)) shows that, at ambient oxygen partial pressure ($\sim$0.21 atm), iron is stable as Fe$_2$O$_3$ (hematite). Consequently, oxide layers formed during sample preparation are expected to persist during permeation testing and act as barriers to hydrogen adsorption and dissociation. This explains why the Fe/Pd and Pd/Fe/Pd configurations do not exhibit measurable permeation despite the presence of a Pd layer. While a single oxide layer (as in Fe/p/Pd) may still allow limited hydrogen transport, potentially through defects exposing the underlying Fe, the presence of oxide layers on both sides introduces a much stronger kinetic barrier that effectively suppresses permeation. In particular, the oxide layer at the Pd/Fe interface limits hydrogen transfer into the substrate, decoupling the enhanced catalytic activity of Pd from the bulk metal.\\

In contrast, acid pickling removes the native oxide layer and enables the attainment of a clean metallic interface prior to Pd deposition, allowing the Pd coating to function effectively as both a catalytic and protective layer. Under these conditions, the surface boundary condition approaches the ideal case assumed in classical permeation analysis, enabling the measurement of intrinsic transport properties. Although complete oxide removal cannot be conclusively confirmed without detailed surface characterisation (e.g., TEM or XPS), which is beyond the scope of this study, the present results demonstrate that minimising surface-controlled limitations requires both oxide removal and catalytic activation. The Pd/p/Fe/p/Pd configuration is therefore identified as the most suitable condition for isolating bulk transport behaviour under the present experimental conditions.

\subsubsection{Influence of hydrogen reduction treatment}

\begin{figure}[ht!]
    \centering
    \includegraphics[width=1\linewidth]{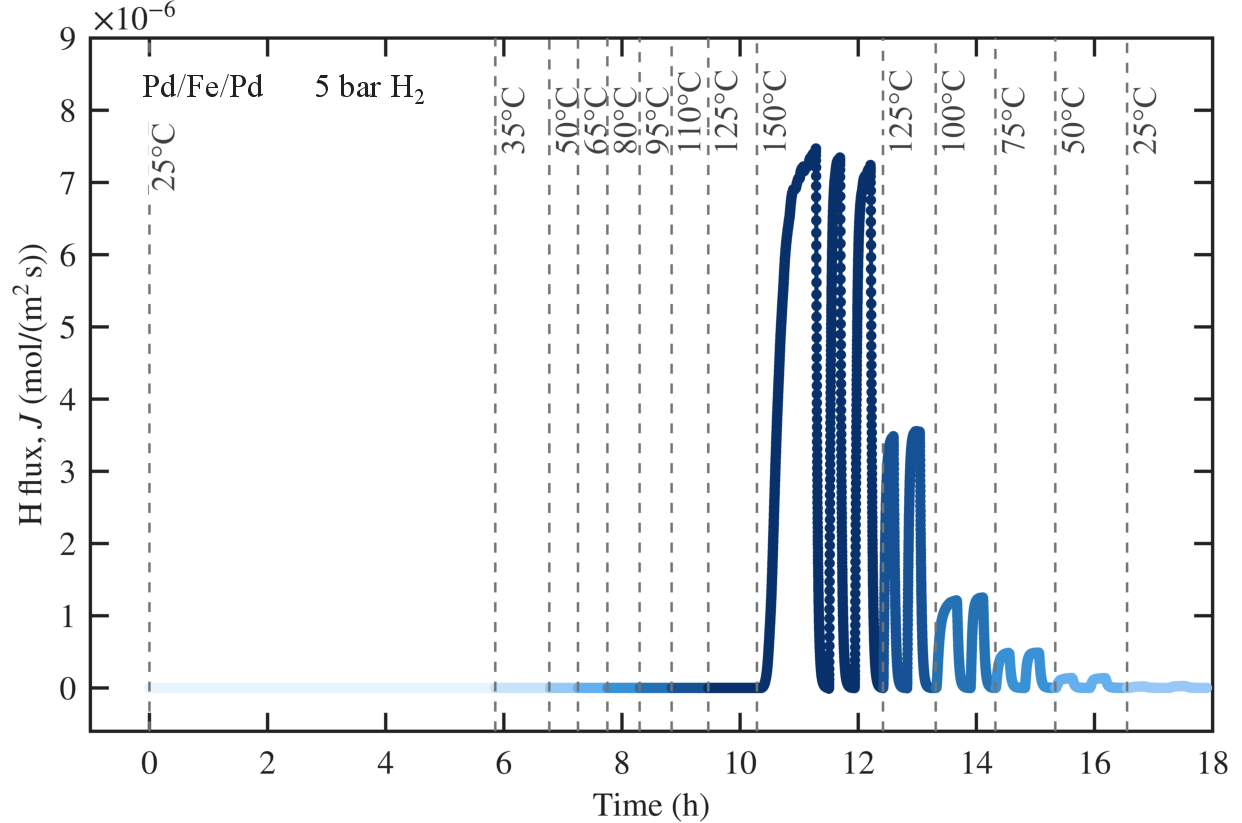}
    \caption{Representative hydrogen reduction experiment for a Pd/Fe/Pd specimen subjected to stepwise heating from 25 to 150 $^\circ$C followed by cooling under 5 bar H$_2$. The signal is background-subtracted; data during heating/cooling ramps and background stabilisation periods are omitted for clarity.}
    \label{Fig7_150C_ActivationProfile}
\end{figure}

Fig. \ref{Fig7_150C_ActivationProfile} shows the measured hydrogen flux as a function of time for a representative hydrogen reduction experiment performed on a sample with the Pd/Fe/Pd surface condition, in which the temperature was progressively increased from 25 to 150 $^\circ$C followed by stepwise cooling back to room temperature under a constant hydrogen pressure of 5 bar.\\ 

Consistent with the negligible permeation previously observed for this surface condition (Fig. \ref{Fig5_PdCoatingComparision}(c)), no measurable permeation signal was detected during the initial room temperature exposure, even after 6 h. This contrasts sharply with oxide-free conditions (Pd/p/Fe/p/Pd), where permeation is generally detected within minutes (Fig. \ref{Fig5_PdCoatingComparision}(d)). Upon increasing temperature, no detectable permeation was observed until reaching 150 $^\circ$C, at which point a strong permeation transient appeared, enabling acquisition of two additional consecutive transients at this temperature. During subsequent cooling, measurable permeation was observed at all evaluated temperatures, including 25 $^\circ$C under identical pressure conditions. This behaviour indicates that hydrogen exposure at 150 $^\circ$C promotes partial reduction of surface and interfacial oxides, thereby restoring hydrogen entry.\\

\begin{figure}[ht!]
    \centering
    \includegraphics[width=1\linewidth]{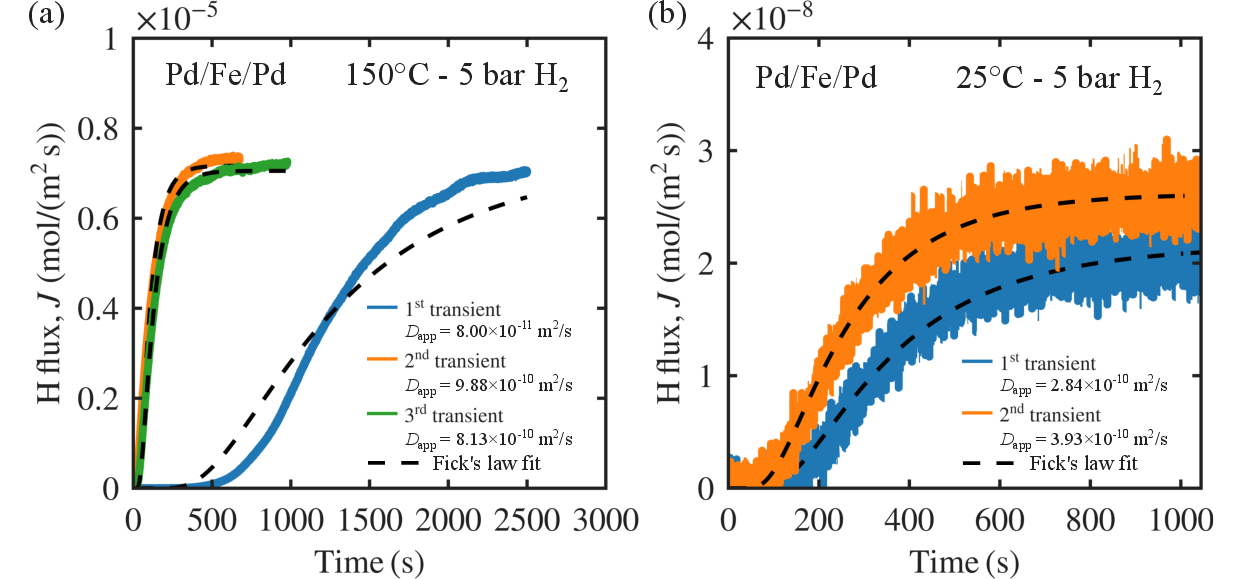}
    \caption{Hydrogen permeation rise transients for a Pd/Fe/Pd sample under 5 bar H$_2$: (a) at 150 $^\circ$C during hydrogen reduction and (b) at 25 $^\circ$C after prior exposure to 150 $^\circ$C. The apparent diffusion coefficients ($D_{\text{app}}$) were determined by fitting the transient permeation curves (dashed lines) to the analytical solution of Fick's law, Eq. (\ref{eq:JtFou}).}
    \label{Fig8_150C_25C_flux}
\end{figure}

Further insight is provided by the transient behaviour shown in Fig. \ref{Fig8_150C_25C_flux}, where the rise transients at 150 $^\circ$C and 25 $^\circ$C are compared. At  150 $^\circ$C, the first transient is significantly delayed relative to subsequent transients, resulting in an approximately one order of magnitude lower $D_{\text{app}}$. The subsequent transients exhibit similar responses and comparable $D_{\text{app}}$ values, indicating that most of the oxide reduction occurs during the first transient. Since trapping effects are expected to be negligible at this temperature \cite{chen_hydrogen_2025}, this evolution reflects progressive surface activation associated with oxide reduction during initial hydrogen exposure. Following this first transient, the permeation response stabilises and reflects intrinsic transport behaviour. Critically, after cooling to 25 $^\circ$C, clear hydrogen permeation transients can be observed, in contrast to the behaviour noted prior to hydrogen reduction. The corresponding $D_{\text{app}}$ values are comparable to those obtained for oxide-free Pd/p/Fe/p/Pd samples, with minor variations likely attributable to experimental scatter and limited trapping effects \cite{kiuchi_solubility_1983,zafra_relative_2023}.\\

\begin{figure}[ht!]
    \centering
    \includegraphics[width=0.9\linewidth]{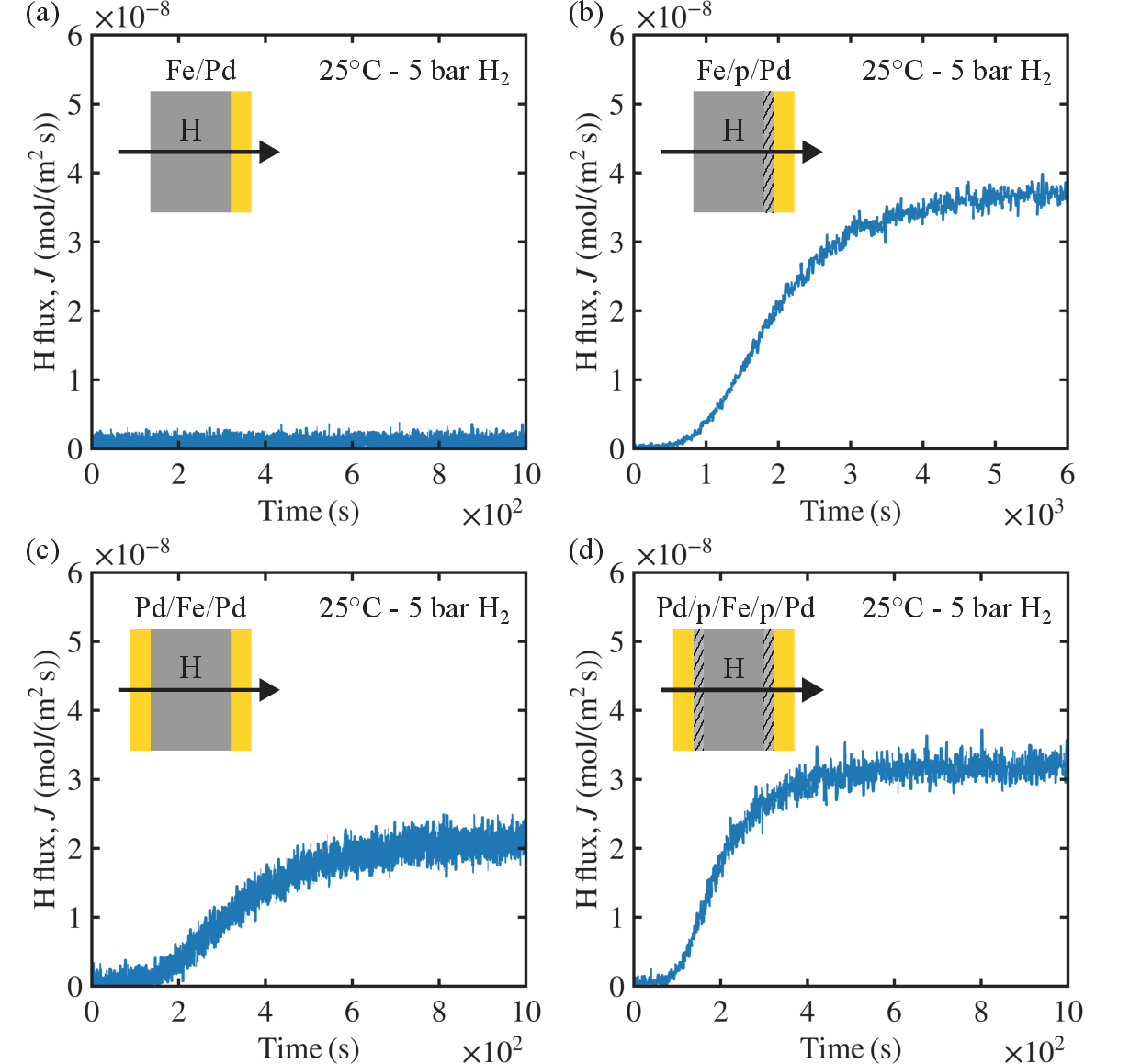}
    \caption{Hydrogen permeation transients measured at 25 $^\circ$C and 5 bar for different sample surface preparations following hydrogen reduction treatment at 150 $^\circ$C: (a) Fe/Pd, (b) Fe/p/Pd, (c) Pd/Fe/Pd, and (d) Pd/p/Fe/p/Pd. Only the first transient is shown.}
    \label{Fig9_CoatCompareAftReduce}
\end{figure}

The permeation transients obtained at 25 $^\circ$C and 5 bar after temperature-enabled hydrogen reduction for each surface condition are shown in Fig. \ref{Fig9_CoatCompareAftReduce}, with the corresponding $D_{\text{app}}$ and $J_{ss}$ values summarised in Table \ref{tab:PermeParameters}. Similar to the results in Fig. \ref{Fig8_150C_25C_flux}(b) for Pd/Fe/Pd (repeated in Fig. \ref{Fig9_CoatCompareAftReduce}(c)), clear permeation transients are also obtained for the Fe/p/Pd and Pd/p/Fe/p/Pd configurations (Figs. \ref{Fig9_CoatCompareAftReduce}(b) and (d)). In contrast, the Fe/Pd condition exhibits no measurable permeation at 25 $^\circ$C after the reduction treatment, consistent with the absence of detectable flux during the reduction stage at 150 $^\circ$C. This behaviour, although not generally expected, suggests that hydrogen reduction at 150 $^\circ$C may not always be sufficient to reduce oxide-covered surfaces completely. Likely, variations in oxide thickness, stoichiometry, and defect structure will significantly influence the reduction treatment efficacy, such that certain oxide layers remain stable under these relatively mild reducing conditions \cite{fradet_thermochemical_2023}. Evaluation of $D_{\text{app}}$ before and after reduction reveals two key trends. First, the reduction treatment leads to an order of magnitude increase in $D_{\text{app}}$ for the Fe/p/Pd and Pd/p/Fe/p/Pd configurations, indicating enhanced hydrogen entry following partial oxide removal. Second, configurations with Pd present on both entry and exit surfaces (Pd/Fe/Pd and Pd/p/Fe/p/Pd) exhibit higher $D_{\text{app}}$ values than Fe/p/Pd, reflecting the beneficial role of Pd in facilitating both hydrogen uptake and subsequent oxide reduction.

\begin{table}[H]
 \centering
  \caption{Summary of apparent diffusivity ($D_{\text{app}}$) and steady-state flux ($J_{ss}$) measured in pure iron at 25 $^\circ$C and 5 bar for different surface conditions, before and after hydrogen reduction treatment at 150 $^\circ$C. Values correspond to the first permeation transient. "NM" indicates that no measurable permeation signal was detected.}
  \label{PermeationFitBC}
\begin{tabular}[t]{cccccc}
\toprule
\multirow{2}{*}{Surface condition} & \multicolumn{2}{c}{Before reduction} & \multicolumn{2}{c}{After reduction}\\
& \text{$D_{\text{app}}$ (m$^2$/s)} & \text{$J_{ss}$} mol/(m$^{2}$s) & \text{$D_{\text{app}}$ (m$^2$/s)} & \text{$J_{ss}$} mol/(m$^{2}$s) \\
\midrule
Fe/Pd & NM & NM & NM & NM\\
\midrule
Fe/p/Pd & 5.46$\pm$0.35$\times10^{-12}$ & 2.84$\pm$2.33$\times10^{-9}$ & 5.39$\times10^{-11}$ & 4.06$\times10^{-8}$\\
\midrule
Pd/Fe/Pd & NM & NM & 2.84$\times10^{-10}$ & 2.16$\times10^{-8}$ \\
\midrule
Pd/p/Fe/p/Pd & 7.38$\pm$0.53$\times10^{-11}$ & 3.01$\pm$0.40$\times10^{-8}$ & 5.19$\times10^{-10}$ & 3.17$\times10^{-8}$\\
\bottomrule
\end{tabular}
\label{tab:PermeParameters}
\end{table}

These results are consistent with the presence of interfacial oxide barriers in unpickled configurations (Fe/Pd, Pd/Fe/Pd), which strongly affect hydrogen permeation at room temperature \cite{ma_influence_2021}. Upon exposure to hydrogen at elevated temperature, these oxide layers can be progressively reduced, restoring hydrogen entry and enabling permeation. The propensity for hydrogen-enabled oxide reduction can be rationalised from a thermodynamic perspective using the equilibrium oxygen potential diagram shown in Fig. \ref{Fig6_EllinghamDiagram}(b). While hydrogen does not readily reduce iron oxides at room temperature, the effective oxygen partial pressure ($p_{O_2}$) in the system is governed by the H$_2$/H$_2$O equilibrium. Increasing temperature and/or lowering the oxygen potential, progressively destabilising iron oxides and favouring metallic Fe. Consequently, oxide layers that are stable at room temperature can become reducible at even moderately elevated temperatures. However, the efficacy of reduction is expected to increase with higher temperatures ($\sim$200-900 $^\circ$C) and depends strongly on oxide morphology and experimental conditions \cite{fradet_thermochemical_2023}. Partial reduction at lower temperatures is therefore limited to thin or defective oxide layers \cite{pineau_kinetics_2006}. In addition, the presence of Pd promotes oxide reduction by catalysing hydrogen dissociation and supplying atomic hydrogen, which can diffuse across the Pd/oxide interface and facilitate oxygen removal \cite{spreitzer_reduction_2019, yarar_pd_2023,oconnor_hydrogen_2020}. As a result, the effective reduction temperature for thin oxide layers is shifted to the $\sim$50-150 $^\circ$C range, although the extent of reduction remains strongly dependent on oxide properties, including oxide thickness, stoichiometry, and defect structure. The present results therefore suggest that this reduction process occurs progressively during hydrogen exposure and is reflected in the evolution of the permeation transients. However, it should be borne in mind that the proposed reduction mechanism is interpreted from the permeation response and thermodynamic considerations, rather than from direct post-test surface-chemical evidence, which would require dedicated surface characterisation.

\subsubsection{Influence of residual gas background}

\begin{figure}[ht!]
    \centering
    \includegraphics[width=1\linewidth]{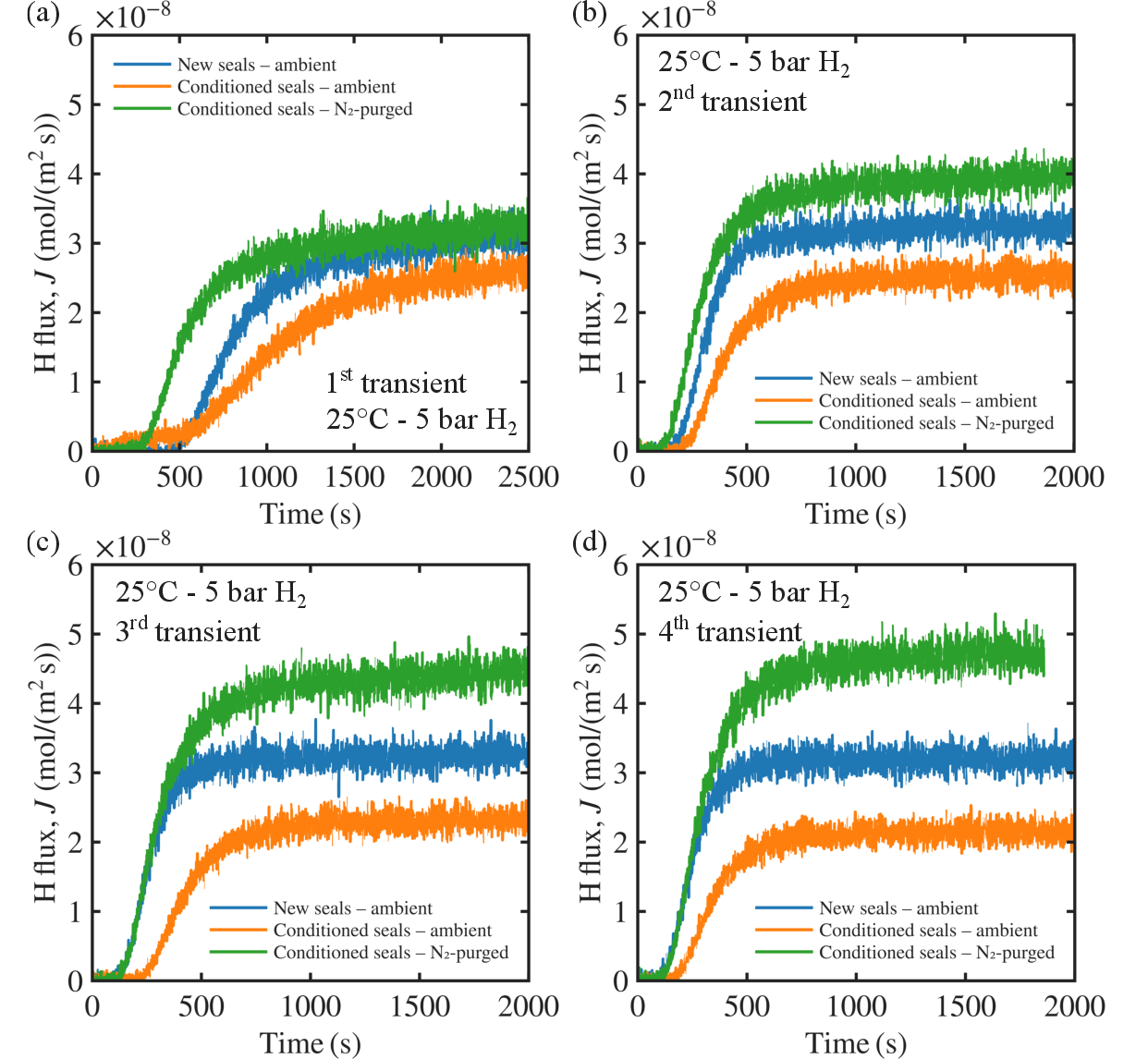}
    \caption{Representative hydrogen permeation rise transients (hydrogen flux vs. time) obtained at 25 $^\circ$C and 5 bar H$_2$ for different sealing and environmental configurations: (a) first transient, (b) second transient, (c) third transient, and (d) fourth transient. “Ambient” denotes operation under laboratory atmosphere, while “N$_2$-purged” refers to the use of a local inert chamber continuously flushed with high-purity nitrogen. “Conditioned seals” correspond to previously used O-rings in high-temperature experiments, whereas “new seals” were used without prior conditioning.}
    \label{Fig10_RTFluxSeal}
\end{figure}

Figure \ref{Fig10_RTFluxSeal} presents the hydrogen permeation rise transients for four consecutive cycles obtained under the three configurations considered to investigate the influence of residual gas: new seals – ambient, conditioned seals – ambient, and conditioned seals – N$_2$-purged. A consistent ranking in hydrogen flux is observed across all transients. The highest flux is systematically obtained under inert purge conditions, despite the use of conditioned seals, followed by the configuration with new seals under ambient conditions, while the lowest flux corresponds to conditioned seals operated under ambient atmosphere. This trend is already apparent in the first transient (Fig. \ref{Fig10_RTFluxSeal}(a)), although the differences between configurations are relatively small at this stage due to the strong influence of hydrogen trapping during the initial charging. As the number of transients increases (Fig. \ref{Fig10_RTFluxSeal}(b)-(d)), the influence of trapping diminishes as trapping sites become progressively occupied \cite{diaz_simulation_2020}, allowing the effect of residual gas background and sealing configuration on hydrogen entry conditions to be more clearly resolved. Accordingly, the separation between the curves becomes more pronounced, particularly between the N$_2$-purged and ambient conditions. In addition, the characteristic time required to reach steady-state flux ($J_{ss}$) remains similar across all configurations, suggesting that sealing and environmental conditions do not significantly affect the overall timescale of hydrogen transport.\\

\begin{figure}[ht!]
    \centering
    \includegraphics[width=1\linewidth]{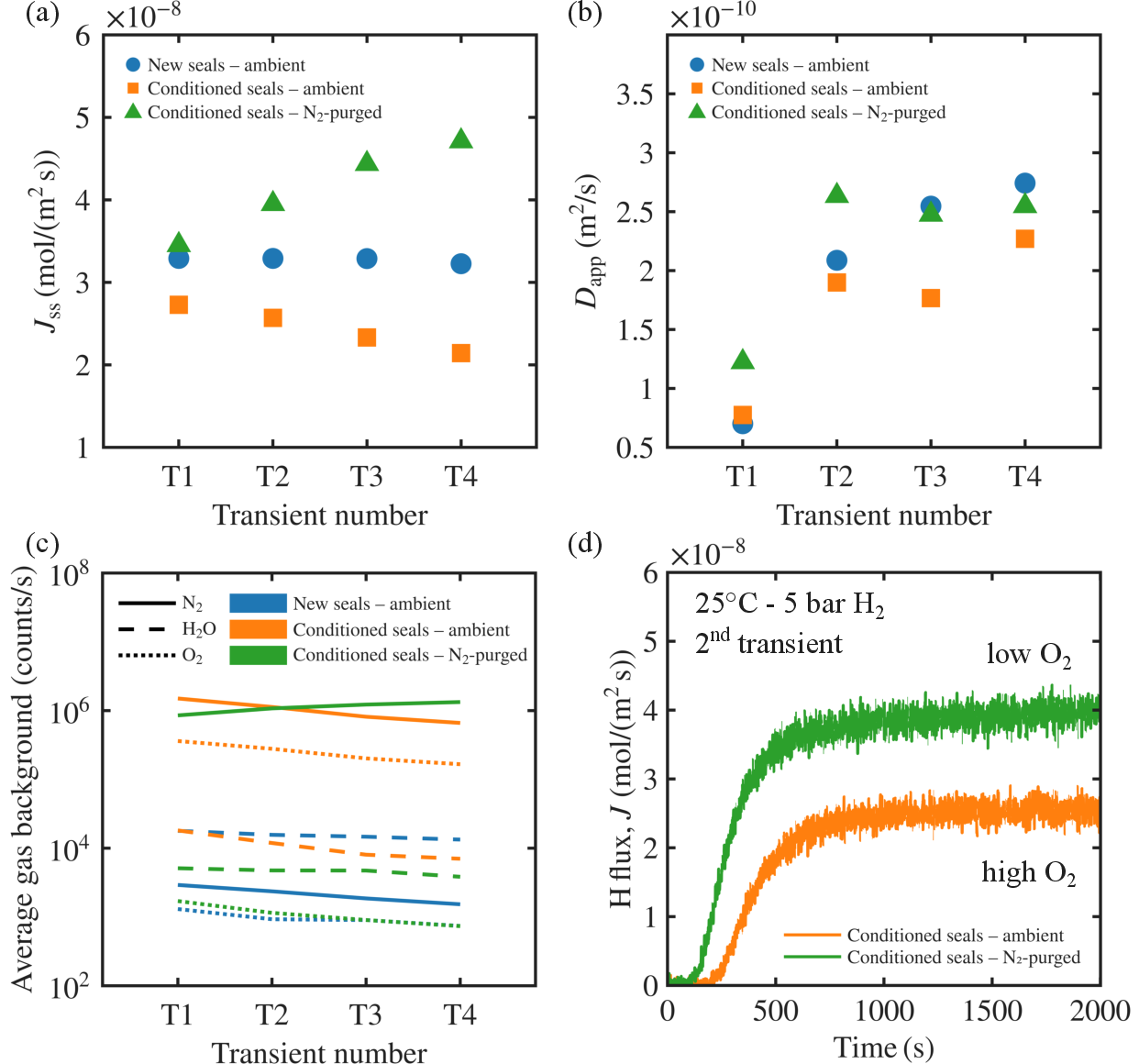}
    \caption{Parameters derived from hydrogen permeation transients obtained at 25 $^\circ$C and 5 bar H$_2$ for different sealing and environmental configurations. (a) Steady-state flux ($J_{ss}$) and (b) apparent diffusivity ($D_{\text{app}}$) as a function of transient number (T1-T4). (c) Average intensity of residual gas species (N$_2$, H$_2$O, and O$_2$) measured during the background period preceding each transient. (d) Representative hydrogen permeation transients (second transient) illustrating the effect of residual O$_2$ levels. “Ambient” refers to operation under laboratory atmosphere, while “N$_2$-purged” denotes the use of a local inert chamber continuously flushed with high-purity nitrogen. “Conditioned seals” correspond to previously used O-rings, whereas “new seals” were used without prior conditioning.}
    \label{Fig11_RTSeal}
\end{figure}

To quantify these trends, $J_{ss}$ and $D_{\text{app}}$ were determined for each transient and are summarised in Fig. \ref{Fig11_RTSeal} as a function of the transient number (T1--T4). The evolution of $J_{ss}$ (Fig. \ref{Fig11_RTSeal}(a)) is shown to be dependent on sealing and environmental conditions. The highest flux is consistently obtained under inert purge conditions, increasing from approximately 3.4 $\times$ 10$^{-8}$ to 4.7 $\times$ 10$^{-8}$ mol/(m$^{2}$s) between the first and fourth transient. In contrast, the lowest flux values are observed for conditioned seals under ambient atmosphere, decreasing slightly from 2.7 $\times$ 10$^{-8}$ to 2.1 $\times$ 10$^{-8}$ mol/(m$^{2}$s). The configuration with new seals under ambient conditions exhibits intermediate behaviour, remaining relatively stable around 3.3 $\times$ 10$^{-8}$ mol/(m$^{2}$s). For both ambient configurations, the steady-state flux remains nearly constant across successive transients. Since steady-state flux is, by definition, independent of trapping (see Eq. (\ref{eq:Jss})), this suggests that gas impurities do not change significantly across transients (i.e., $C_L^0$ is relatively independent of the transient number). However, this is not the case for the conditioned seals N$_2$ purged condition, where $J_{ss}$ increases with transient number, likely due to the drop in O$_2$ with transient number, shown in Fig. \ref{Fig11_RTSeal}(c). This also highlights the relatively higher importance of O$_2$ relative to other impurities.\\ 

The evolution of $D_{\text{app}}$ (Fig. \ref{Fig11_RTSeal}(b)) reflects the influence of trapping in addition to sealing and environmental effects. For all configurations, $D_{\text{app}}$ increases significantly from the first to the second transient, which is attributed to the initial presence of empty trapping sites that reduce the effective hydrogen mobility during the first charging step. As these traps become progressively occupied, their influence diminishes, and from the second transient onwards, $D_{\text{app}}$ stabilises around 2-3 $\times$ 10$^{-10}$ m$^{2}$/s. Differences between configurations are comparatively small, confirming that bulk hydrogen transport is only weakly affected by the residual gas background.\\ 

The different flux behaviours observed across configurations can be rationalised by inspecting the evolution of the residual gas background, as shown in Fig. \ref{Fig11_RTSeal}(c). The highest O$_2$ levels are consistently measured for conditioned seals under ambient atmosphere, whereas the lowest O$_2$ levels are obtained under inert purge conditions and when new seals are used in ambient conditions. In contrast, N$_2$ levels are highest under inert purge conditions and for conditioned seals exposed to ambient atmospheres, while H$_2$O shows comparatively smaller variations between configurations. Elevated O$_2$ signals for the conditioned, ambient atmosphere case are consistent with minor air ingress through degraded elastomeric O-ring seals during evacuation \cite{kommlingInsightsLifetimePredictions2020}, as supported by the significant reductions in O$_2$ levels when operating under inert purge conditions and when using new seals. When considered in terms of average residual gas background, a direct correspondence can be observed between O$_2$ level and steady-state flux: lower O$_2$ background is associated with higher $J_{ss}$, whereas higher O$_2$ background leads to reduced flux. This behaviour is consistent across all configurations and transients and indicates that the residual gas background primarily affects hydrogen entry conditions rather than bulk transport, as further supported by the comparatively weak variation in $D_{\text{app}}$.\\ 

The dependence of $J_{ss}$ on O$_2$ level suggests that oxygen-containing species present before hydrogen admission modify the effective surface boundary condition for hydrogen entry, most likely through adsorption or partial poisoning of the Pd surface, thereby reducing the efficiency of hydrogen dissociation and absorption. Conversely, under lower residual gas background conditions, the Pd surface remains more catalytically active, facilitating hydrogen uptake and resulting in higher steady-state flux. This effect is further illustrated in Fig. \ref{Fig11_RTSeal}(d), where a direct comparison between two configurations with comparable N$_2$ levels and similar H$_2$O content but distinct O$_2$ background shows that lower O$_2$ levels lead to a significantly higher permeation flux. The progressive increase in flux observed under inert purge conditions, despite a measured O$_2$ background comparable to that obtained using new seals under ambient conditions, further indicates that improved environmental control during evacuation leads to a progressively cleaner and more favourable surface state for hydrogen uptake between transients. This is consistent with the similar $J_{ss}$ values measured during the first transient for both configurations, and suggests that N$_2$ purging may improve hydrogen uptake beyond what can be inferred from the measured O$_2$ signal alone, analogous to purging procedures commonly employed in hydrogen mechanical testing. However, the progressive increase in $J_{ss}$ also indicates that the effective boundary condition is still evolving under the N$_2$-purged configuration. Therefore, this condition is mainly useful for identifying the influence of environmental control and residual gas background, whereas the new-seal ambient configuration provides a more stable and reproducible condition for diffusion characterisation in the present study. If N$_2$ purging were continued for additional cycles, the flux would be expected to eventually stabilise once the surface/environmental state reaches a steady condition, although the precise mechanisms responsible for this effect will be the subject of future work.\\

Overall, these results demonstrate that relatively small variations in residual gas background -- particularly O$_2$ -- governed by sealing condition and environmental exposure during evacuation, can lead to systematic and reproducible differences in measured hydrogen permeation behaviour -- especially steady-state hydrogen flux -- even when all other experimental parameters and surface conditions are nominally identical. Notably, the magnitude of these variations is comparable to the experimental scatter commonly reported in hydrogen permeation studies, suggesting that uncontrolled residual gas background may represent a significant and often overlooked source of variability across different setups. It is important to emphasise that these experiments do not correspond to controlled H$_2$/O$_2$ gas mixtures, but rather to variations in residual O$_2$ background prior to H$_2$ admission. Once the HPS is pressurised with high-purity hydrogen, the system operates under positive pressure, preventing further air ingress, and any residual gas present under vacuum is expected to be strongly diluted. Therefore, the observed effect cannot be systematically attributed to a sustained oxygen partial pressure in the hydrogen gas phase, but instead arises from pre-existing surface conditions established during evacuation. This distinction highlights the difference between pre-adsorbed oxygen effects and those arising from intentionally introduced H$_2$/O$_2$ gas mixtures. Under an initially high O$_2$ background, the Pd-coated entry surface is likely modified by the adsorption of oxygen-containing species prior to hydrogen exposure. Although the Pd layer prevents oxidation of the underlying iron substrate, it remains susceptible to surface poisoning or modification, which can reduce the efficiency of hydrogen dissociation and absorption. This introduces an additional activation step for hydrogen uptake, leading to a reduced effective hydrogen flux. In contrast, under a lower residual O$_2$ background, the Pd surface remains more catalytically active, resulting in enhanced hydrogen entry and higher steady-state flux.\\

These findings are directly relevant to the broader discussion on the role of gas impurities in hydrogen permeation. Recent studies on controlled H$_2$/impurity mixtures report contrasting behaviours: while R\"othig et al. \cite{rothig_gaseous_2025} observed minimal and mostly transient effects of O$_2$ (100 ppm-12000 ppm), Zhou et al. \cite{zhou_impact_2024} reported a strong inhibitory effect, including significant reductions in steady-state permeation and diffusivity at relatively low O$_2$ concentrations (350 ppm). Although the present results are qualitative in this context, they demonstrate that even trace residual contamination prior to hydrogen charging can substantially influence steady-state flux, while diffusivity remains comparatively unaffected. This highlights the strong sensitivity of hydrogen entry to surface condition and system cleanliness, and underscores the importance of monitoring background and controlling background gas levels in permeation experiments. The residual O$_2$ signal therefore provides a practical metric of experimental quality and a key parameter for reliable interpretation and comparison of hydrogen permeation data.

\subsubsection{Influence of pressure and temperature}

\begin{figure}[ht!]
    \centering
    \includegraphics[width=1\linewidth]{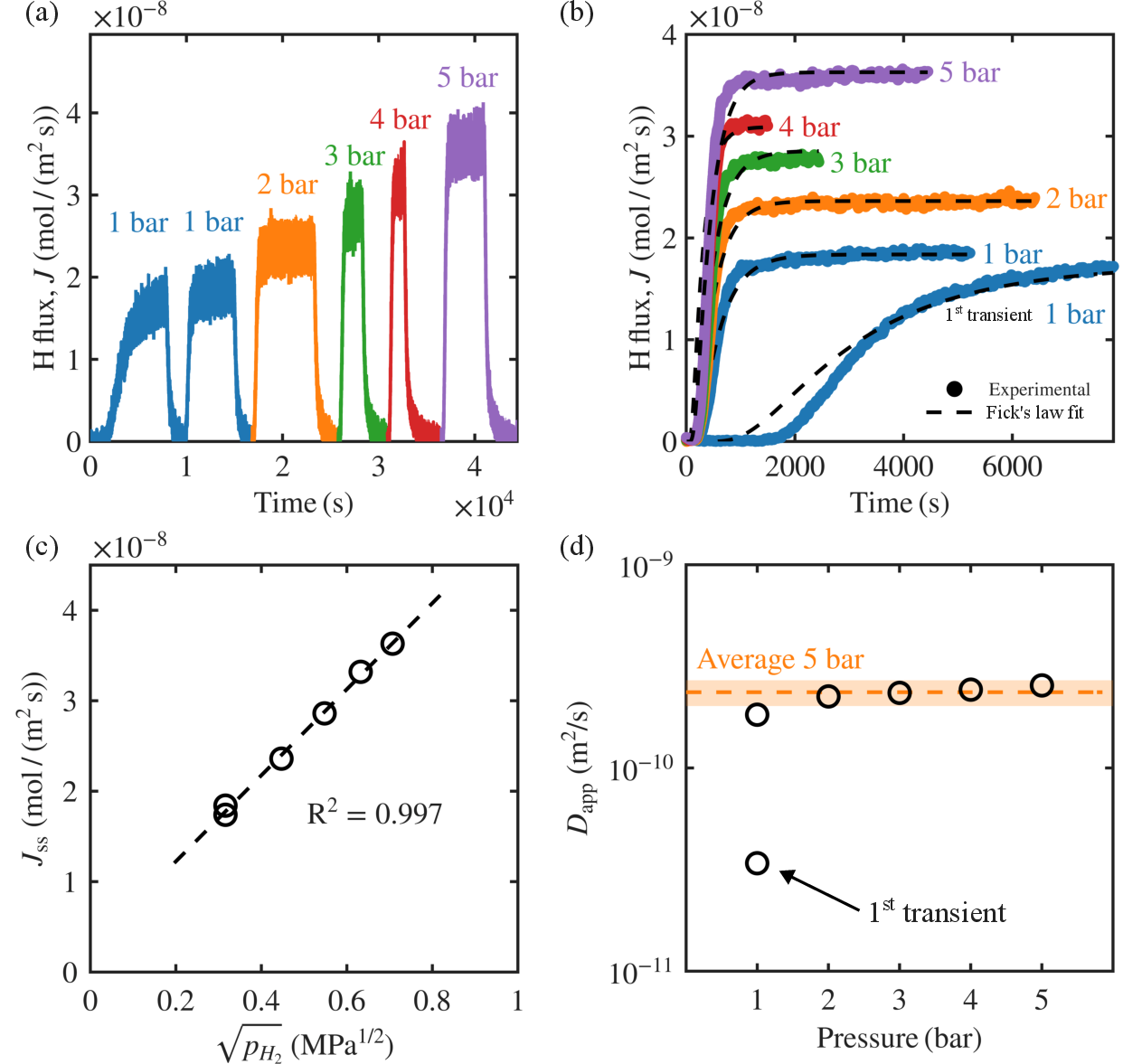}
    \caption{Pressure dependence of hydrogen permeation in pure iron at 25 $^\circ$C for the Pd/p/Fe/p/Pd surface condition (membrane thickness $L$=0.8 mm). (a) Evolution of hydrogen flux during a stepwise pressurisation experiment (1-5 bar), including repeated transients at 1 bar. (b) Comparison of the corresponding rise transients for each pressure. (c) Steady-state flux ($J_{ss}$) as a function of $\sqrt{p_{H_2}}$, showing a linear dependence consistent with Sieverts’ law. (d) Apparent diffusivity ($D_{\text{app}}$) as a function of pressure, illustrating the strong effect of trapping during the first transient and the stabilisation of diffusivity at higher pressures.}
    \label{Fig12_FluxPressureDependence}
\end{figure}

After establishing an optimised surface condition that minimises the influence of surface oxides and associated artefacts (Pd/p/Fe/p/Pd), the effect of applied hydrogen pressure on permeation behaviour at 25 $^\circ$C was investigated. Fig. \ref{Fig12_FluxPressureDependence}(a) shows the evolution of hydrogen flux during a stepwise pressurisation experiment, in which successive rise-decay transients were performed between 1 and 5 bar. Two transients were conducted at 1 bar to minimise the influence of trapping. For clarity, the corresponding rise transients are compared in Fig. \ref{Fig12_FluxPressureDependence}(b), where a systematic increase in steady-state flux with increasing hydrogen pressure is clearly observed. Although the first transient at 1 bar exhibits a delayed response due to the presence of initially empty trapping sites, it ultimately converges to a $J_{ss}$ value comparable to that of the second transient at 1 bar, where trapping effects are significantly reduced, and steady-state is reached more rapidly. This result confirms the theoretical deduction that $J_{ss}$ is independent of trapping.\\

The relationship between steady-state flux and hydrogen pressure is further analysed in Fig. \ref{Fig12_FluxPressureDependence}(c), which shows $J_{ss}$ as a function of $\sqrt{p_{H_2}}$. A strong linear correlation (R$^2$=0.997) is obtained, demonstrating that the permeation behaviour follows Sieverts' law; i.e., that, considering Eqs. (\ref{eq:Sievert}) and (\ref{eq:Jss}), the steady-state flux can be related to the H$_2$ pressure as,
\begin{equation}
  J_{{ss}}=\frac{D_L S_L}{L}\sqrt{p_{H_2}}
\end{equation}
\noindent and, for low trap occupancy,
\begin{equation}
   J_{{ss}}=\frac{D_{\text{app}} S_{\text{app}}}{L}\sqrt{p_{H_2}} 
\end{equation}
\noindent as per Eq. (\ref{eq:AppEqs}). The linear dependence observed confirms that hydrogen transport is governed by bulk lattice diffusion under the present experimental conditions, with negligible influence from surface kinetic limitations. This is consistent with the optimised surface preparation, which minimises oxide-related barriers and enables near-ideal boundary conditions for hydrogen entry. The linear fit exhibits a small positive intercept of 1.65 $\times$ 10$^{-9}$ mol H/(m$^2$ s) at zero pressure, which closely matches the independently determined instrumental LOD of 1.98 $\times$ 10$^{-9}$ mol H/(m$^2$ s). This agreement provides additional confidence in the quantitative reliability of the measurements and confirms that the observed permeation signals are well above the detection threshold.\\

The evolution of $D_{\text{app}}$ with pressure is shown in Fig. \ref{Fig12_FluxPressureDependence}(d). A pronounced increase is observed between the first and second transients at 1 bar, from 4.5 $\times$ 10$^{-11}$ to 1.8 $\times$ $10^{-10}$ m$^2$/s, which is attributed to the progressive filling of trapping sites during the initial charging step. Beyond this initial transient, $D_{\text{app}}$ exhibits only a slight increase with pressure and rapidly stabilises around 2.4 $\times$ 10$^{-10}$ m$^2$/s, consistent with the values obtained in previous experiments at 5 bar. This weak pressure dependence suggests that in pure iron with low trap density, the influence of pressure on trapping is minimal once traps are saturated. Within experimental scatter, a consistent diffusivity is therefore obtained for pressures above 1 bar, confirming that the measured transport parameters are representative of intrinsic bulk diffusion.\\

\begin{figure}[ht!]
    \centering
    \includegraphics[width=1\linewidth]{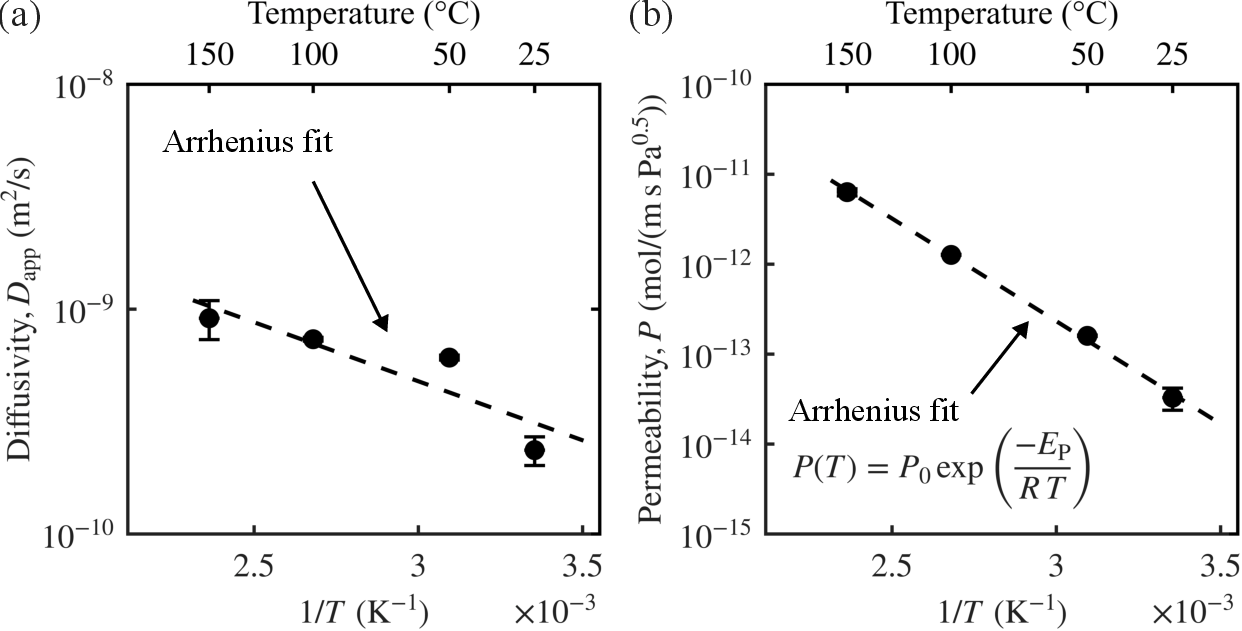}
    \caption{Temperature sensitivity of hydrogen transport parameters in pure iron for the Pd/p/Fe/p/Pd surface condition: (a) measured (apparent) diffusivity ($D_{\text{app}}$), and (b) permeability ($P$). Data correspond to the mean of the last three permeation transients at each temperature (25 - 150 $^\circ$C), with the first transient excluded to minimise trapping effects. Error bars indicate standard deviation. Dashed lines show linear Arrhenius fits used to extract pre-exponential factors and activation energies.}
    \label{Fig13_ArrheniusPlot}
\end{figure}

In addition to pressure, temperature is a key thermodynamic variable governing hydrogen transport in metals. Figs. \ref{Fig13_ArrheniusPlot}(a) and (b) show the temperature dependence of the apparent diffusivity ($D_{\text{app}}$) and permeability ($P$), respectively, presented as Arrhenius plots (log-linear scale) as a function of inverse temperature. The data were obtained for samples with the optimised surface condition (Pd/p/Fe/p/Pd) and correspond to the average values of the last three permeation transients conducted at each temperature (25, 50, 100 and 150 $^\circ$C), with the first transient excluded to minimise the influence of hydrogen trapping. Error bars represent the standard deviation at each temperature.\\ 

Both $D_{\text{app}}$ and $P$ (which is determined from $J_{ss}$ using Eq. (\ref{eq:Perme})) increase with temperature, revealing the expected influence of temperature on the diffusion and solubility of hydrogen. The associated activation energies can be obtained by fitting the data to an Arrhenius law; e.g., for the permeability:
\begin{equation}\label{eq:Arrhenius}
  P=P_0\exp\left(-\frac{E_P}{RT}\right)
\end{equation}

\noindent where $P_0$ is the pre-exponential factor, $E_P$ is the activation energy, $R$ is the universal gas constant, and $T$ is the absolute temperature. It should be noted that, in theory, the apparent diffusivity should not follow an Arrhenius law; only the lattice diffusivity does. However, when the measured diffusivity approaches $D_L$, its variation with respect to temperature can be well approximated by an Arrhenius law. This is indicated by the dashed line in Fig. \ref{Fig13_ArrheniusPlot}(a); it can be seen that the agreement is satisfactory at high temperatures, where trapping effects are negligible. The fit to the $D_{\text{app}}$ vs $T$ data delivers a pre-exponential factor of $D_0 = 1.8$ $\times$ 10$^{-8}$ m$^2$/s, close to the values associated with lattice diffusivity \cite{garcia2024tds}, and an activation energy of $E_D=10.05$ kJ/mol. For the case of permeability, the Arrhenius fit delivers values of $P_0=1.65$ $\times$ 10$^{-6}$ mol/(m s Pa$^{0.5}$) and $E_P=43.71$ kJ/mol, which are within the range of values reported in the literature \cite{bryan_diffusivity_1963}, and reflect the combined contributions of diffusion and hydrogen dissolution, as permeability inherently incorporates both mobility and solubility effects. The variation with temperature of the permeability, an inherently lattice quantity, is very accurately described by an Arrhenius law, demonstrating that our measurements can successfully remove surface effects and operate in the bulk-controlled regime.

\subsection{Discussion}
\subsubsection{Summary of the results}

\begin{figure}[H]
    \centering
    \includegraphics[width=0.9\linewidth]{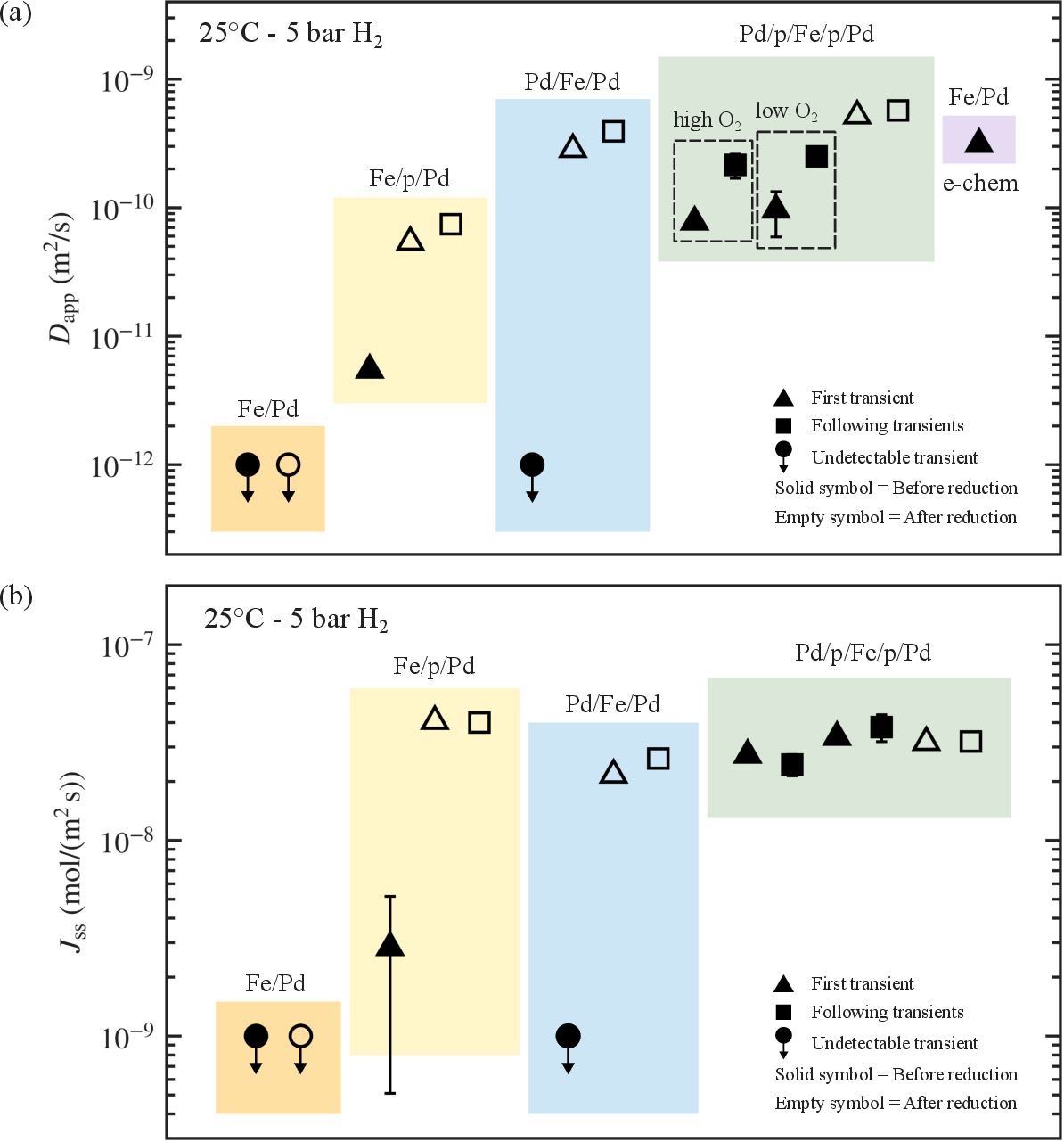}
    \caption{Summary of (a) apparent diffusivity ($D_{\text{app}}$) and (b) steady-state flux ($J_{ss}$) obtained under the different experimental conditions investigated in this work. Results are shown for 25 $^\circ$C. Each point corresponds to either a single permeation transient or an average value obtained from repeated or consecutive transients, as appropriate, with the first transient indicated separately. The data highlight the influence of surface preparation, residual gas background (low/high O$_2$), and hydrogen reduction treatment on the measured permeation response.}
    \label{Fig14_Summary_25C}
\end{figure}

\begin{figure}[H]
    \centering
    \includegraphics[width=1\linewidth]{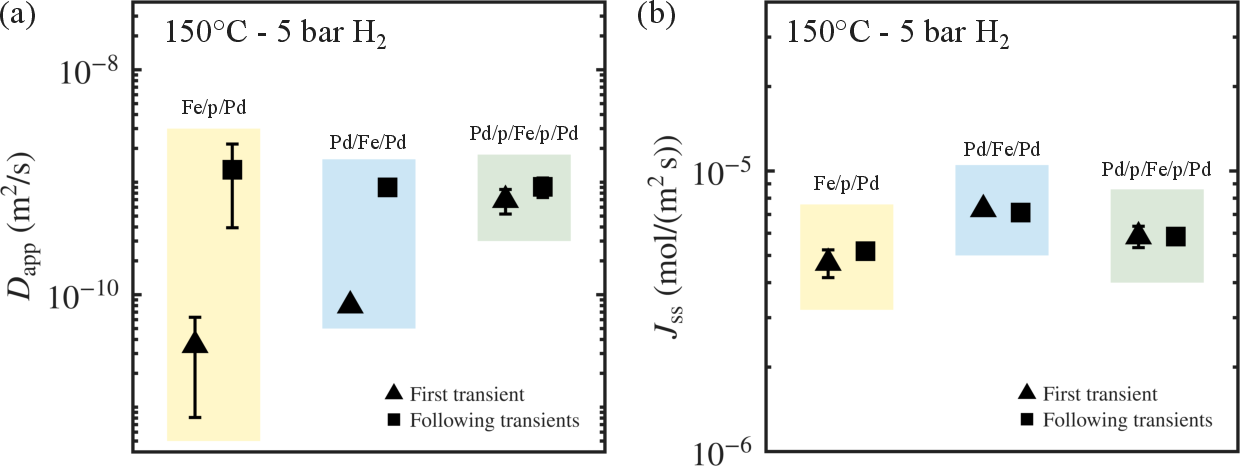}
    \caption{Summary of (a) apparent diffusivity ($D_{\text{app}}$) and (b) steady-state flux ($J_{ss}$) obtained under the different experimental conditions investigated in this work. Results are shown for 150 $^\circ$C. Each point corresponds to either a single permeation transient or an average value obtained from repeated or consecutive transients, as appropriate, with the first transient indicated separately. The data highlight the influence of surface preparation and hydrogen reduction treatment on the measured permeation response.}
    \label{Fig15_Summary_150C}
\end{figure}

The apparent diffusivity ($D_{\text{app}}$) and steady-state flux ($J_{ss}$) obtained for the different experimental conditions investigated in this work, which spanned variations in surface preparation, residual gas background (oxygen level), and hydrogen reduction treatment, are summarised in Figs. \ref{Fig14_Summary_25C} and  \ref{Fig15_Summary_150C}. Results are shown for both 25 $^\circ$C and 150 $^\circ$C, enabling direct comparison of the relative contributions of surface-controlled and bulk-controlled processes, respectively. Furthermore, a distinction is made between the first permeation transient and subsequent transients for each condition, as the initial measurement is influenced by transient effects (e.g., surface activation and trapping), whereas subsequent transients are more representative of steady-state transport behaviour.

At 25 $^\circ$C (Fig. \ref{Fig14_Summary_25C}), considerable differences in $D_{\text{app}}$ ($\sim$100-fold, when measurable) and $J_{ss}$ ($\sim$50-fold) are observed depending on the surface condition and testing environment. The lowest values correspond to configurations where a native oxide layer is present on the HPS at either the metal surface or the Pd/metal interface (e.g., Fe/Pd, Fe/p/Pd and Pd/Fe/Pd before reduction). In such cases, measurable permeation is either not detected (Fe/Pd and Pd/Fe/Pd, indicated by the downward black arrows in Fig. \ref{Fig14_Summary_25C}) or only observed after prolonged exposure times (Fe/p/Pd). This results in concomitant reductions in both $D_{\text{app}}$ and $J_{ss}$ compared to the nominally oxide-free Pd/p/Fe/p/Pd condition as well as traditional electrochemical permeation data, which is included for reference in Fig. \ref{Fig14_Summary_25C}(a). Closer evaluation of these pre-reduction treatment results provides important mechanistic insights into the role of native oxides on hydrogen uptake and diffusion. For example, both Fe/Pd and Fe/p/Pd are expected to have a native oxide present on the HPS, while only Fe/Pd would have an oxide layer on the LPS. Critically, the delayed, but clear permeation transient for Fe/p/Pd indicates that the HPS oxide layer was unable to completely mitigate hydrogen uptake. Such a result implies that the oxide layer either is sufficiently defected to enable slow, but tangible hydrogen uptake directly at the Fe substrate or only provides a moderate barrier to hydrogen adsorption, dissociation, and diffusion. The absence of measurable permeation for the Pd/Fe/Pd and Fe/Pd conditions further suggests that the additional impedance of a second native oxide at the LPS, when combined with the effects of the HPS oxide, is sufficient to effectively prevent hydrogen transport over the evaluated test durations. Conversely, the nominal absence of an oxide layer on both the HPS and LPS of the Pd/p/Fe/p/Pd condition results in rapid permeation.\\

Intermediate $D_{\text{app}}$ values are obtained for partially conditioned systems, including configurations exposed to a high residual gas background (labelled as ``high O$_2$''), where surface contamination or partial oxidation of the Pd layer reduces the effective hydrogen entry rate. The Fe/p/Pd configuration after hydrogen reduction treatment also falls within this intermediate range, suggesting only partial reduction of surface oxides and, consequently, partial recovery of hydrogen transport relative to nominally oxide-free conditions. This postulated incomplete reduction is likely due to insufficient time at 150 $^\circ$C \cite{pineau_kinetics_2006}. In contrast, higher $D_{\text{app}}$ values at 25 $^\circ$C are consistently obtained for the Pd/p/Fe/p/Pd condition under low residual gas background (excluding first transient), where surface and trapping effects are minimised. The highest $D_{\text{app}}$ values correspond to Pd/Fe/Pd and Pd/p/Fe/p/Pd configurations after reduction treatment, indicating that the presence of Pd promotes more effective oxide reduction \cite{spreitzer_reduction_2019, yarar_pd_2023,oconnor_hydrogen_2020}. This is attributed to the catalytic role of Pd in dissociating molecular hydrogen and supplying atomic hydrogen at the surface, which enhances hydrogen uptake and thus facilitates the reduction of oxide species at the metal interface. As a result, this activation step remains beneficial even in systems where oxides are expected to have been largely removed (Pd/p/Fe/p/Pd). Although the trends observed for $D_{\text{app}}$ are broadly similar to those obtained for $J_{ss}$, with the highest fluxes corresponding to well-conditioned systems, an important exception is observed for Fe/p/Pd, for which $J_{ss}$ lies within the upper range despite the comparatively low $D_{\text{app}}$. This suggests that oxide layers have a stronger influence on transient behaviour than on steady-state flux. Overall, excluding configurations where oxide layers remain unmodified (i.e., without pickling or reduction), the variability in both $D_{\text{app}}$ and $J_{ss}$ is limited and falls within the scatter commonly reported in the literature for hydrogen transport in bcc iron \cite{zafra_relative_2023}. For example, the diffusivity obtained from electrochemical permeation measurements (first transient) on the same material is also included and shows good agreement with the present gas permeation results under conditions where oxide effects are minimised. This is consistent with the use of cathodic polarisation during the electrochemical testing, which suppresses oxide formation on the analogue of the HPS for gas permeation and thus facilitates hydrogen uptake \cite{liu_determination_2014}. \\

The trends observed at 25 $^\circ$C can be rationalised by considering the behaviour at 150 $^\circ$C, where hydrogen-induced reduction of surface oxides becomes significant. As shown in Figs. \ref{Fig15_Summary_150C}(a) and (b), except for Fe/Pd (which was undetectable and hence not plotted in the figure), all surface configurations exhibit measurable hydrogen permeation at this temperature, including those that were inactive at room temperature (e.g., Fe/p/Pd and Pd/Fe/Pd), as hydrogen-enabled reduction/activation of surface and interfacial oxides at elevated temperature weakens the kinetic barrier for hydrogen entry \cite{pineau_kinetics_2006}. The absence of measurable permeation for the Fe/Pd condition suggests that hydrogen reduction at 150 $^\circ$C may not always be sufficient to activate oxide-covered surfaces, likely due to variations in oxide thickness, stoichiometry, and defect structure, which can significantly influence reducibility. A distinctive feature at 150 $^\circ$C is the behaviour of the first transient, which exhibits significantly lower $D_{\text{app}}$ values (by up to two orders of magnitude) compared to subsequent transients. Unlike at room temperature, where this effect is partly associated with hydrogen trapping, the negligible trapping expected at elevated temperatures \cite{chen_hydrogen_2025} instead suggests that permeation is being affected by surface activation processes, likely related to the progressive reduction of residual surface oxide during initial hydrogen exposure. Following this initial transient, the permeation response stabilises and reflects the intrinsic transport properties of the material. Consequently, from the second transient onwards, both $D_{\text{app}}$ and $J_{ss}$ converge towards a narrow range characteristic of bulk diffusion control, largely independent of the initial surface condition. Consistent with this, $J_{ss}$ remains comparatively less affected by surface oxides and interfacial effects, as it is determined under steady-state conditions after sufficient time for surface processes to stabilise and, by definition, is independent of trapping (see Eq. (\ref{eq:Jss})); accordingly, variations in $J_{ss}$ are primarily governed by surface phenomena that influence the effective lattice hydrogen concentration at the subsurface ($C_L^0$), whereas $D_{\text{app}}$ -- derived from the transient response -- remains more sensitive to early-stage surface limitations.\\

In summary, these results establish the significant influence that native oxides have on the hydrogen permeation behaviour of pure iron. For example, changes in surface preparation, the use of elevated temperature hydrogen reduction treatments, and the presence of different residual gas backgrounds were capable of inducing order-of-magnitude differences in observed hydrogen diffusivity. Such findings emphasise the importance of controlling and reporting surface preparation and environmental conditions when interpreting and comparing hydrogen permeation data. However, what remains unclear is how these native oxides mechanistically alter hydrogen permeation, particularly with respect to understanding the relative contribution of the oxide as a diffusion barrier versus its effects on hydrogen uptake processes. Future efforts should focus on executing carefully controlled experiments with different oxide thicknesses and types of oxides to deconvolute these two possible contributions. Such understanding could then enable the development of thermal pretreatments to induce preferred oxides to hinder hydrogen permeation in critical applications.

\subsubsection{Comparison with the literature}
\begin{figure}[H]
    \centering
    \includegraphics[width=1\linewidth]{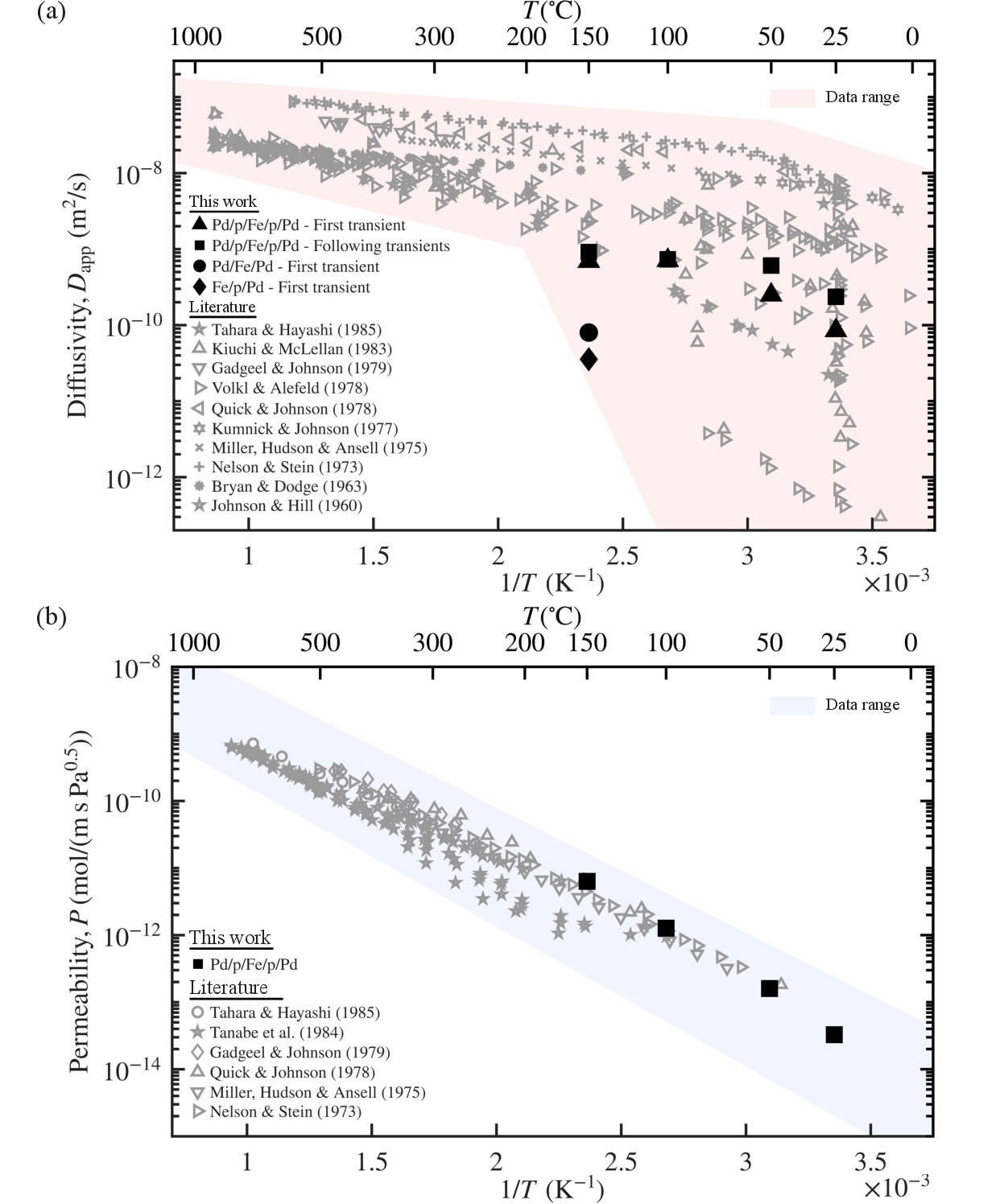}
    \caption{(a) Hydrogen diffusivity ($D_{\text{app}}$) and (b) permeability ($P$) in pure Fe as a function of inverse temperature. Literature data for diffusivity were taken from Refs.~\cite{taharaMeasurementsPermeationHydrogen1985,gadgeel_gas-phase_1979,quickHydrogenDeuteriumIron1978,kumnickHydrogenDeuteriumIron1977,miller_permeation_1975,Nelson1973,bryan_diffusivity_1963,johnsonDiffusivityHydrogenAlpha1960}, while permeability data were taken from Refs.~\cite{tanabe_hydrogen_1984,quickHydrogenDeuteriumIron1978,miller_permeation_1975,gadgeel_gas-phase_1979,Nelson1973,taharaMeasurementsPermeationHydrogen1985}. The datasets compiled in Refs ~\cite{volklDiffusionHydrogenMetals1978,kiuchi_solubility_1983} are reviews, and the reader is referred to the original sources therein. Results from the present work are included for comparison. The shaded regions indicate the overall scatter of reported values in the literature. Error bars are omitted, as the associated uncertainty is smaller than the symbol size.}
    \label{Fig16_CompareLit}
\end{figure}

The results of this study provide an opportunity to comment on potential mechanistic contributors to previously reported trends and observations regarding hydrogen-metal interactions in pure iron from the literature. Towards this end, a compilation of gas permeation-based measurements of hydrogen diffusivity and permeability data as a function of temperature (spanning 0 to 1000 $^\circ$C) from both literature and the current study is presented in Figs. \ref{Fig16_CompareLit}(a) and (b), respectively. \\

As has been broadly discussed in the literature \cite{o.d.gonzalezPARTVICommunications1968,kiuchi_solubility_1983}, the hydrogen diffusivity exhibits (1) substantial variation amongst authors, particularly at lower temperatures, and (2) a notable inflection to a steeper slope occurs as the temperature is reduced below 200 $^\circ$C. Conversely, the hydrogen permeability data for pure iron exhibits significantly reduced scatter compared to the diffusivity, as well as a generally linear relationship between inverse temperature and the logarithm of the permeability over all evaluated temperatures.\\

Considering first the hydrogen permeability, Fig. \ref{Fig16_CompareLit}(b), its variation as a function of temperature is consistent with the literature on pure iron. For example, fitting the data from the present study to an Arrhenius relationship (Eq. (\ref{eq:Arrhenius})) yields an activation energy of permeation ($E_{P}$) of 43.7 kJ/mol, which lies within the range of values reported in the literature (30-50 kJ/mol) \cite{Nelson1973, bryan_diffusivity_1963,choi_diffusion_1970}. This reduced scatter relative to the hydrogen diffusivity of iron can be attributed to the fact that permeability is evaluated under steady-state conditions. As such, hydrogen permeability measurements are expected to be less sensitive to surface influences or other transient effects than the hydrogen diffusivity, which is explicitly evaluated using transient data. Critically, this intrinsic difference in variability implies that the uncertainty in hydrogen solubility measurements (derived by the ratio of permeability to diffusivity) will be strongly influenced by the accuracy of the measured diffusivity \cite{SanMarchi2012}.\\

Similar to the hydrogen permeability, all hydrogen diffusivities measured in the current study are contained within the envelope of the compiled literature data as a function of temperature, as demonstrated in Fig. \ref{Fig16_CompareLit}(a). Critically, the differences in surface conditions used in the current study can be leveraged to understand the respective contributions of hydrogen trapping and surface effects to the observed variability in literature diffusivity data. For example, considering the optimal Pd/p/Fe/p/Pd configuration, comparison of the diffusivity determined from sequential transients at 25 $^\circ$C provides information on the influence of trapping. As shown in Fig. \ref{Fig14_Summary_25C}, variations in diffusivity were observed between the first and later transients for the Pd/p/Fe/p/Pd condition, but such differences were generally within a factor of three. Such a 3-fold difference strongly suggests that trapping effects cannot explain the substantial scatter -- spanning nearly five orders of magnitude -- in hydrogen diffusivity observed for pure iron at 25 $^\circ$C. This postulation is consistent with the findings of Hagi et al. \cite{hagi_diffusion_1979}, who performed electrochemical permeation experiments on both single-crystal and polycrystalline iron. Critically, these authors reported that the hydrogen diffusivity was only weakly affected by the presence of grain boundaries, which are expected to be the dominant trap in well-annealed, pure Fe, similarly indicating that bulk trapping alone is insufficient to explain the observed variability in literature data.\\ 

However, the surface condition used during the testing and, by extension, the presence of native oxides appear to have a significant influence on the measured hydrogen diffusivity. For example, a nearly 30-fold difference in $D_{\text{app}}$ was observed between the first and later transients conducted on the Fe/p/Pd condition at 150 $^\circ$C (Fig. \ref{Fig15_Summary_150C}). Similar differences between first and subsequent transients were also noted for the Pd/Fe/Pd condition. Given the expectation that oxide reduction will occur during the first transient and thus not be present (or at least significantly degraded) for subsequent permeation transients, these results imply that the initially reduced diffusivity is likely due to kinetic limitations from the native oxide. In other words, the current results indicate that surface effects, particularly oxides, strongly contribute to the observed permeation behaviour, with the effect expected to increase in magnitude as temperature decreases. This inference provides a possible mechanistic explanation for the well-documented 'knee' in hydrogen diffusivity for pure iron, which is typically characterised by a break in slope between approximately 150 and 200 $^\circ$C (as shown in Fig. \ref{Fig16_CompareLit}(a)). Moreover, such surface and oxide effects may also explain the observed scatter in hydrogen diffusivity, particularly at low temperatures. For example, differences in sample preparation methods (e.g., use of pickling) or testing methodologies (e.g., performing pre-treatments at higher temperature) would be expected to induce order of magnitude changes in measured diffusivity, as shown in Table \ref{tab:PermeParameters}. It should be noted that the present study is limited to temperatures up to 150 $^\circ$C, and that higher-temperature experiments, which are currently ongoing and will be the subject of future work, are required to draw definitive conclusions on this point.\\

Critical examination of the literature generally supports a governing contribution of surface effects to the observed scatter and 'knee' in hydrogen diffusivity for pure iron as a function of temperature. Considering the influence on the observed scatter in measurements, Miller et al. \cite{miller_permeation_1975} reported that untreated specimens or Pd-coated samples without appropriate activation exhibited strongly surface-controlled behaviour, particularly at low temperature, thereby preventing reliable measurements below 342 K. In contrast, studies employing aggressive pre-treatments, such as prolonged hydrogen exposure at high temperature, tend to report higher and more consistent diffusivity values. Gonzalez \cite{o.d.gonzalezPARTVICommunications1968}, for instance, exposed membranes to hydrogen for 72 h at 615 $^\circ$C before testing, while Nelson and Stein \cite{Nelson1973} demonstrated that reproducible results required pre-exposure to hydrogen at 500–600 $^\circ$C to reduce surface oxides. Similarly, Kumnick and Johnson \cite{kumnickHydrogenDeuteriumIron1977} showed using a gas permeation system with electrochemical detection that reliable measurements could only be obtained using Pd-coated specimens on both entry and exit surfaces, effectively eliminating surface impedance effects. Even studies that report apparently linear Arrhenius behaviour can still exhibit subtle curvature when examined closely. For example, Quick and Johnson \cite{quickHydrogenDeuteriumIron1978} reported a single Arrhenius relationship for diffusivity, even though a noticeable inflection occurs in their data at approximately 200 $^\circ$C. This distinction is important when comparing these literature datasets with the present results, since the diffusivity values reported here correspond to specimens subjected only to controlled surface preparation, without any prior high-temperature hydrogen reduction or thermal activation treatment. Therefore, the relative position of the present data within the literature envelope should be interpreted in the context of differences in pre-treatment history, Fe purity, membrane thickness, gas pressure and other methodological variables that vary between studies.\\ 

The potential for modifications in hydrogen-surface interactions to significantly alter hydrogen permeation is fully consistent with the broader literature on iron and steels. For example, permeation experiments by Bruzzoni and Garavaglia \cite{AnodicIronOxide1992a} demonstrated that anodically formed oxide films introduce an additional resistance to hydrogen transport. This `barrier' effect was also observed in recent studies showing that compact oxide layers can act as effective barriers to hydrogen permeation \cite{zhang_effects_2018}. Similarly, atom probe tomography revealed that oxide layers on steel can act as hydrogen trapping sites, delaying but not completely suppressing hydrogen ingress \cite{wang_atom_2025}. Moreover, Kiuchi and McLellan \cite{kiuchi_effect_1983} explicitly highlighted that uncontrolled surface conditions have historically compromised hydrogen transport measurements in pure iron due to this 'barrier' effect, advocating rigorous surface preparation as a prerequisite for reliable data. Considering these observations in the context of the observed 'knee' in diffusivity, it appears likely that this inflection is due to surface effects introducing a kinetic barrier to hydrogen entry and thus a reduced diffusivity at lower temperatures. Such an interpretation is supported by literature observations and the present results. For example, Bryan and Dodge \cite{bryan_diffusivity_1963} did not observe an inflection in diffusivity during testing on pure iron. This is likely due to their employed minimum testing temperature ($\sim$126 $^\circ$C), which was sufficiently elevated to avoid the low-temperature regime where surface effects become dominant. Specifically, as temperature increases, hydrogen-assisted oxide reduction progressively removes this barrier, resulting in a transition to bulk-controlled diffusion and recovery of the intrinsic Arrhenius behaviour. This effect is consistent with the result shown in Fig. \ref{Fig7_150C_ActivationProfile}, where permeation was not observed until a temperature of 150 $^\circ$C was reached. Interestingly, this `onset' temperature is closely aligned with the temperature where the inflection in diffusivity is generally observed ($\sim$150–200 $^\circ$C). Such strong correlation further suggests that the diffusivity `knee' is likely an artefact arising from surface-controlled transport.\\

In summary, the present results indicate that the observed scatter and associated 'knee' in hydrogen diffusivity for pure iron, particularly at low temperatures, likely originates from uncontrolled variations in surface condition and oxide state, rather than from intrinsic bulk phenomena such as trapping. As such, surface-controlled effects can significantly affect the permeation response, even under nominally identical testing conditions, leading to order-of-magnitude discrepancies between studies. These findings emphasise the critical importance of carefully controlling, characterising, and reporting surface preparation and environmental conditions in hydrogen permeation experiments. While the influence of surface condition has been recognised for decades, the present study advances this understanding by systematically separating the contributions of pickling, Pd coating, residual gas background and hydrogen reduction/activation treatment to the measured gaseous permeation response. In doing so, it provides a practical framework for distinguishing surface-controlled artefacts from bulk transport behaviour. More broadly, these results highlight the need to decouple surface and bulk contributions in order to establish reliable intrinsic hydrogen transport properties, an issue that remains insufficiently resolved in the literature despite its fundamental and technological relevance. We note in passing that the optimised surface conditions used in this work are not intended to reproduce a specific service environment, but to provide a controlled reference state in which oxide-related surface limitations are minimised and the intrinsic hydrogen transport response of the metal can be isolated. More realistic service conditions, including native oxides, coatings, mechanical loading, gas impurities and hydrogen flow, can then be systematically introduced in future gaseous permeation studies.

\section{Conclusions}
\label{Sec:ConcludingRemarks}

In this work, a high-sensitivity mass spectrometry-based gaseous hydrogen permeation system has been developed and validated. The system operates at pressures up to 50 bar and temperatures up to 250 $^\circ$ C, with straightforward scalability to more extreme conditions. A key feature is its high detection sensitivity, achieving a hydrogen flux limit of $1.98 \times 10^{-9}$~mol/(m$^{2}$~s), which enables reliable quantification of very low permeation fluxes and accurate resolution of transient permeation behaviour. In addition, a comprehensive framework for the analysis of hydrogen permeation, including the governing transport equations and their practical interpretation, is presented. The application of this system to a model material (annealed pure Fe) enabled a systematic investigation of the influence of surface preparation and key testing variables. The main findings are summarised as follows:

\begin{itemize}
\item Hydrogen permeation at room temperature is strongly governed by surface-controlled processes. The presence of native oxide layers can severely limit or completely suppress hydrogen uptake, even when Pd coatings are applied. When oxide layers are present on both entry and exit surfaces, permeation is effectively blocked, whereas the presence of a single oxide layer allows limited permeation, albeit with strongly hindered kinetics. Reliable and reproducible permeation measurements require removal of surface oxides through pickling and catalytic activation via Pd coating on both sides of the membrane. Under these conditions, the surface boundary approaches the ideal case assumed in classical permeation theory, enabling measurement of intrinsic transport properties.

\item Hydrogen exposure at elevated temperature promotes progressive reduction of surface and interfacial oxides, restoring hydrogen entry and enabling permeation in previously inactive configurations. This process is captured directly through the evolution of successive permeation transients, with the first transient reflecting surface activation. However, oxide reduction at 150 $^\circ$C is not universally effective, as demonstrated by the Fe/Pd condition, indicating that oxide reducibility depends strongly on oxide thickness, stoichiometry, and defect structure. These results highlight that oxide layers are not equivalent and that surface activation is inherently system-dependent. 

\item Residual gas contamination, even at trace levels, has a measurable and systematic effect on hydrogen permeation. In particular, O$_2$ present prior to hydrogen exposure modifies the effective surface boundary condition, most likely through adsorption or poisoning of the Pd surface, thereby reducing hydrogen dissociation efficiency and steady-state flux. In contrast, bulk transport, as reflected by the apparent diffusivity, is comparatively insensitive to residual gas background. These findings demonstrate that variations in sealing condition and environmental control can lead to reproducible differences in permeation behaviour and may constitute a significant source of variability across experimental studies. 

\item Under optimised surface conditions, hydrogen permeation in pure iron follows bulk diffusion-controlled behaviour. The steady-state flux exhibits a linear dependence on the square root of hydrogen pressure, consistent with Sieverts’ law, confirming that hydrogen transport is governed by lattice diffusion. The apparent diffusivity shows minimal dependence on pressure once trapping sites are saturated. Both diffusivity and permeability exhibit Arrhenius behaviour with temperature, yielding activation energies consistent with literature values. These results validate the accuracy of the developed system and confirm that intrinsic transport properties can be reliably measured when surface effects are minimised. 

\item The results demonstrate that variations in surface condition, residual gas background, and thermal history can lead to order-of-magnitude differences in measured hydrogen permeation behaviour, even for the same material. In particular, surface effects -- especially oxide-related barriers -- are shown to be a dominant source of variability at low temperatures. This provides a mechanistic explanation for the large scatter and the commonly reported “knee” in hydrogen diffusivity for pure Fe, suggesting that these features arise primarily from surface-controlled phenomena rather than intrinsic bulk properties. Consequently, many discrepancies in the literature can be rationalised by differences in surface preparation and experimental protocols.

\end{itemize}

Overall, the developed high-sensitivity gas permeation system provides a robust and versatile platform for quantitative investigation of hydrogen transport under service-relevant conditions. The present results demonstrate that accurate and reproducible permeation measurements require stringent control of surface state and environmental conditions, particularly with respect to oxide formation and residual gas contamination. Beyond pure Fe, the methodology established in this work provides a foundation for systematic studies of more complex materials and environments, including alloys, coatings, multi-material systems, and controlled impurity effects. Future work will focus on quantitatively decoupling surface and bulk contributions, with particular emphasis on the role of oxide layers and their evolution under realistic service conditions, as well as on advanced surface characterisation to better resolve their impact on hydrogen uptake and permeation behaviour.

\section*{Acknowledgements}
\label{Sec:Acknowledgeoffunding}

\noindent The authors acknowledge financial support from the EPSRC (grant EP/V009680/2) and from the UKRI Horizon Europe Guarantee programme (ERC Starting Grant \textit{ResistHfracture}, EP/Y037219/1). E.\ Mart\'{\i}nez-Pa\~neda additionally acknowledges financial support from UKRI's Future Leaders Fellowship programme [grants MR/V024124/1 and MR/V024124/2]. Z. Harris acknowledges the support of the Richard King Mellon Faculty Fellowship in Energy Research from the University of Pittsburgh.

\section*{Declarations}
\label{Sec:Declarations}

\noindent\textbf{Ethical Approval} This article does not contain any studies involving human participants or animals carried out by any of the authors.

\noindent\textbf{Informed Consent} Not applicable.

\noindent\textbf{Competing Interests} The authors declare that they have no competing interests.

\appendix
\setcounter{table}{0} 
\setcounter{figure}{0} 
\renewcommand\thetable{\Alph{section}.\arabic{table}}  
\renewcommand\thefigure{\Alph{section}.\arabic{figure}} 

\section{Derivation of apparent variables in hydrogen permeation}
\label{Appendix:Apparent}

We proceed to show that, under conditions of low trap and lattice occupancy, the trapped hydrogen concentration is proportional to the lattice one, and consequently, an apparent diffusion coefficient can be defined, and measured, which is independent of concentration and incorporates lattice and trapping contributions. As in the main text, we define a concentration of hydrogen in lattice sites $C_L$ and a concentration of hydrogen in trapping sites $C_T$, which are related to the site density and coverage, such that $C_L=N_L \theta_L$ and $C_T=N_T \theta_T$. 
Oriani's equilibrium \cite{oriani1970diffusion} establishes a relationship between the coverage of lattice ($\theta_L$) and trapping ($\theta_T$) sites,
\begin{equation}
    \frac{\theta_T}{1-\theta_T} = \frac{\theta_L}{1-\theta_L} K \,\,\,\,\,\,\,\, \text{with} \,\,\,\,\,\,\,\, K=\exp \left(\frac{E_b} {RT}\right)
\end{equation}

\noindent where $K$ is an equilibrium constant, defined in terms of the binding energy $E_b$, taking here as a positive constant, the absolute temperature $T$, and the gas constant $R$. Assuming a low occupancy of trapping and lattice sites, $\theta_T << 1$ and $\theta_L << 1$, Oriani's equilibrium can be formulated as,
\begin{equation}
    C_T= \frac{N_T}{N_L}  K C_L
\end{equation}

\noindent Accordingly, the total hydrogen content reads,
\begin{equation}
    C = C_L + C_T = \left( 1 + \frac{N_T}{N_L}  K \right) C_L
\end{equation}

\noindent Hence, the flux equation, Eq. (\ref{eq:Flux}), can be rewritten as,
\begin{equation}
    J = - D_L \frac{\partial C_L}{\partial x} = - D_L \frac{\partial}{\partial x} \left( \frac{C}{1 + \frac{N_T}{N_L}  K} \right) = - \frac{D_L}{1 + \frac{N_T}{N_L}  K} \frac{\partial C}{\partial x} = -D_{\text{app}} \frac{\partial C}{\partial x}
\end{equation}

\noindent such that hydrogen transport can be expressed as a function of the total hydrogen concentration, through an apparent diffusivity $D_{\text{app}}=D_L/(1+N_TK/N_L)$. Similarly, Sieverts' law, Eq. (\ref{eq:Sievert}), can be reformulated in terms of an apparent solubility $S_{\text{app}}$,
\begin{equation}
    S_L \sqrt{p_{H_2}} = 
    C_L^0 = \frac{C^0}{1 + \frac{N_T}{N_L}  K} \implies C^0 = \left( 1 + \frac{N_T}{N_L}  K \right) S_L \sqrt{p_{H_2}} = S_{\text{app}} \sqrt{p_{H_2}}
\end{equation}

\noindent The steady-state flux can also be reformulated in terms of apparent quantities,
\begin{equation}
   J_{ss}=\frac{D_L C_L^0}{L} = \left( 1 + \frac{N_T}{N_L} K \right) D_{\text{app}}\frac{C^0}{1 + \frac{N_T}{N_L} K} = \frac{D_{\text{app}}C^{0}}{L}
\end{equation}

\noindent Finally, by definition, the permeability does not distinguish between apparent and lattice:
\begin{equation}
    P_{\text{app}} = D_{\text{app}} S_{\text{app}} =\frac{D_L}{1 + \frac{N_T}{N_L}  K} S_L \left( 1 + \frac{N_T}{N_L}  K \right)   = D_L S_L = P 
\end{equation}

One should note that, while less relevant to permeation, the hydrogen experimental literature often refers to a diffusible or apparent hydrogen concentration, which is not covered by the reduction to apparent variables above. This is, for example, relevant to isothermal TDS experiments \cite{zafra_comparison_2022,zafra_relative_2023}, where the hydrogen released at room temperature is integrated over a fixed time, such that,
\begin{equation}
    C_{\text{diff}} = \int_0^{t_{\text{exp}}} J_{\text{out}} (t) \, \text{d} t
\end{equation}

\noindent where $J_{\text{out}}$ is the measured egress flux in the isothermal TDS experiment and $t_{\text{exp}}$ the duration of the experiment. While hydrogen can only leave the sample \textit{via} lattice transport, this measured hydrogen concentration encompasses not only the hydrogen residing in lattice sites at the beginning of the experiment but also the hydrogen atoms sequestered in weak traps that have `detrapped' over the time scale of the experiment. Under the assumption of Oriani's equilibrium, detrapping is much faster than diffusion, and the measured hydrogen content should be the total one (assuming $C_L \approx 0$ in vacuum). However, Orani's equilibrium assumption can break down in the presence of strong traps, which can bring the detrapping time scale to be on the order (or larger) than the diffusion time scale of a given experiment. Under this scenario, the measured hydrogen content is not the total one (as Oriani or the low trap occupancy assumption considered above would assume) but a `diffusible' quantity encompassing the lattice hydrogen and that trapped in traps that are sufficiently weak to detrap for the temperature and time of the experiment. Traps with sufficiently slow detrapping kinetics do not contribute to the measured signal and therefore fall outside this experimentally accessible population. In such cases, the assumption of local equilibrium underpinning Oriani-type formulations is no longer valid, and analysis requires a full diffusion–trapping kinetic framework, such as those based on McNabb and Foster's model \cite{martinez-paneda_generalised_2020}.

\setcounter{table}{0} 
\setcounter{figure}{0} 
\renewcommand\thetable{\Alph{section}.\arabic{table}}  
\renewcommand\thefigure{\Alph{section}.\arabic{figure}} 

\section{Additional experiments}
\label{Appendix:B}

This Appendix presents permeation results from two duplicate experiments conducted at 25 $^\circ$C and 5 bar H$_2$ for a surface preparation consisting of Pd/p/Fe. In this configuration, the pickled and Pd-coated surface is exposed to hydrogen on the charging side, while hydrogen is detected on the opposite uncoated Fe surface. A very rapid breakthrough is observed, occurring within a few minutes, together with a comparatively high hydrogen flux that is reached shortly after hydrogen exposure. However, a well-defined steady-state flux ($J_{ss}$) is not established over the duration of the experiments. This indicates that the presence of a native oxide layer on the detection side does not significantly limit hydrogen entry or bulk transport, but instead affects the recombination kinetics of hydrogen at the exit surface, thereby preventing stabilisation of the permeation flux.

\begin{figure}[H]
    \centering
    \includegraphics[width=1\linewidth]{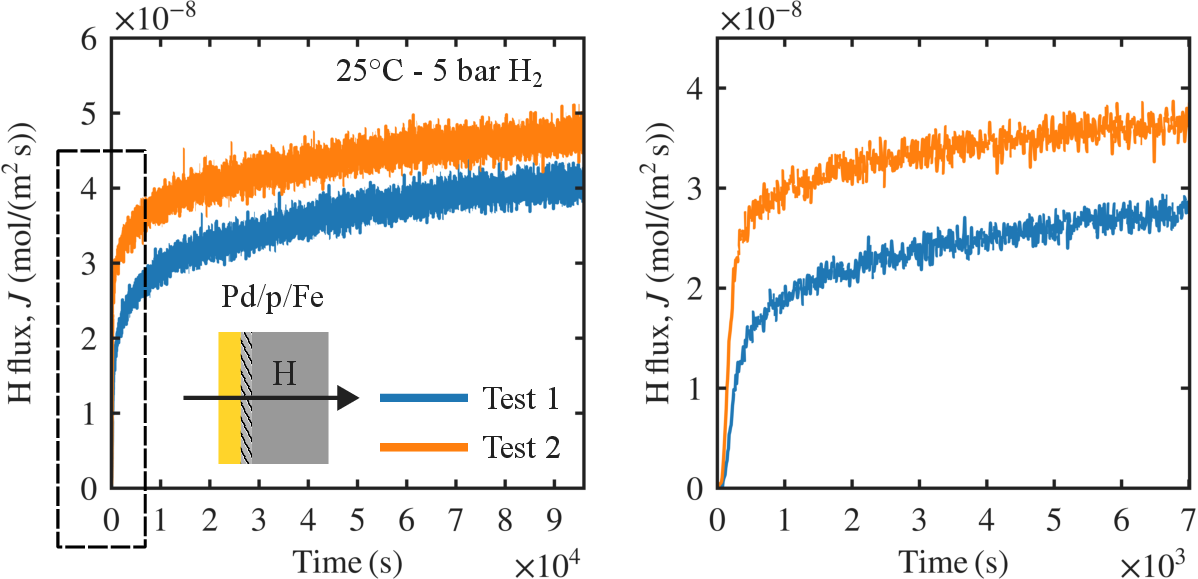}
    \caption{Duplicate hydrogen permeation transients measured at 25 $^\circ$C and 5 bar H$_2$ for the Pd/p/Fe surface preparation (duplicate experiments). Hydrogen enters through the pickled and Pd-coated charging side and is detected on the opposite uncoated Fe surface. A rapid breakthrough is observed, but a steady-state flux is not established over the duration of the experiments, indicating recombination-limited behaviour at the oxidised detection surface. The zoomed region is indicated by the square frame.}
    \label{FigB1_PdpFe}
\end{figure}


\small
\begin{thebibliography}{10}
\expandafter\ifx\csname url\endcsname\relax
  \def\url#1{\texttt{#1}}\fi
\expandafter\ifx\csname urlprefix\endcsname\relax\def\urlprefix{URL }\fi
\expandafter\ifx\csname href\endcsname\relax
  \def\href#1#2{#2} \def\path#1{#1}\fi

\bibitem{gunathilake_comprehensive_2024}
C.~Gunathilake, I.~Soliman, D.~Panthi, P.~Tandler, O.~Fatani, N.~A.
  Ghulamullah, D.~Marasinghe, M.~Farhath, T.~Madhujith, K.~Conrad, Y.~Du,
  M.~Jaroniec, A comprehensive review on hydrogen production, storage, and
  applications, Chemical Society Reviews 53~(22) (2024) 10900--10969.

\bibitem{chen_hydrogen_2025}
Y.-S. Chen, C.~Huang, P.-Y. Liu, H.-W. Yen, R.~Niu, P.~Burr, K.~L. Moore,
  E.~Martínez-Pañeda, A.~Atrens, J.~M. Cairney, Hydrogen trapping and
  embrittlement in metals – {A} review, International Journal of Hydrogen
  Energy 136 (2025) 789--821.

\bibitem{delrio2025local}
F.~W. DelRio, M.~Schmitz-Elbers, U.~Strohmeier, T.~Straub, Local failure strain
  and reduction of area provide early metrics for hydrogen embrittlement in
  microscale austenitic stainless steel tensile specimens, Experimental
  Mechanics (2025) 1--9.

\bibitem{malheiros2022local}
L.~C. Malheiros, A.~Oudriss, S.~Cohendoz, J.~Bouhattate, F.~Th{\'e}bault,
  M.~Piette, X.~Feaugas, Local fracture criterion for quasi-cleavage
  hydrogen-assisted cracking of tempered martensitic steels, Materials Science
  and Engineering: A 847 (2022) 143213.

\bibitem{fernandez-sousa_analysis_2020}
R.~Fernández-Sousa, C.~Betegón, E.~Martínez-Pañeda, Analysis of the
  influence of microstructural traps on hydrogen assisted fatigue, Acta
  Materialia 199 (2020) 253--263.

\bibitem{martinez2016strain}
E.~Mart{\'\i}nez-Pa{\~n}eda, C.~F. Niordson, R.~P. Gangloff, Strain gradient
  plasticity-based modeling of hydrogen environment assisted cracking, Acta
  Materialia 117 (2016) 321--332.

\bibitem{guedes2020role}
D.~Guedes, L.~C. Malheiros, A.~Oudriss, S.~Cohendoz, J.~Bouhattate, J.~Creus,
  F.~Th{\'e}bault, M.~Piette, X.~Feaugas, The role of plasticity and hydrogen
  flux in the fracture of a tempered martensitic steel: A new design of
  mechanical test until fracture to separate the influence of mobile from
  deeply trapped hydrogen, Acta Materialia 186 (2020) 133--148.

\bibitem{noauthor_iso_2014}
{ISO} 17081:2014 – {Method} of measurement of hydrogen permeation and
  determination of hydrogen uptake and transport in metals by an
  electrochemical technique, International {Standard} ISO 17081:2014,
  International Organization for Standardization, Geneva (2014).

\bibitem{noauthor_standard_2018}
Standard {Practice} for {Evaluation} of {Hydrogen} {Uptake}, {Permeation}, and
  {Transport} in {Metals} by an {Electrochemical} {Technique}, International
  {Standard} ASTM G148-97(2018), ASTM International, West Conshohocken, PA, USA
  (2018).

\bibitem{cupertino2024suitability}
L.~Cupertino-Malheiros, T.~K. Mandal, F.~Th{\'e}bault,
  E.~Mart{\'\i}nez-Pa{\~n}eda, On the suitability of single-edge notch tension
  (sent) testing for assessing hydrogen-assisted cracking susceptibility,
  Engineering Failure Analysis 162 (2024) 108360.

\bibitem{bao2026new}
Y.~Bao, Z.~Ouyang, G.~Kerherve, D.~J. Payne, L.~Cupertino-Malheiros, New
  insights into the hydrogen evolution reaction kinetics using advanced
  electrochemical techniques, Electrochimica Acta (2026) 148253.

\bibitem{zafra_comparison_2022}
A.~Zafra, Z.~Harris, C.~Sun, E.~Martínez-Pañeda, Comparison of hydrogen
  diffusivities measured by electrochemical permeation and
  temperature-programmed desorption in cold-rolled pure iron, Journal of
  Natural Gas Science and Engineering 98 (2022) 104365.

\bibitem{hageman2022electro}
T.~Hageman, E.~Mart{\'\i}nez-Pa{\~n}eda, An electro-chemo-mechanical framework
  for predicting hydrogen uptake in metals due to aqueous electrolytes,
  Corrosion Science 208 (2022) 110681.

\bibitem{cupertino2024hydrogen}
L.~Cupertino-Malheiros, M.~Duportal, T.~Hageman, A.~Zafra,
  E.~Mart{\'\i}nez-Pa{\~n}eda, Hydrogen uptake kinetics of cathodic polarized
  metals in aqueous electrolytes, Corrosion Science 231 (2024) 111959.

\bibitem{liu_determination_2014}
Q.~Liu, A.~D. Atrens, Z.~Shi, K.~Verbeken, A.~Atrens, Determination of the
  hydrogen fugacity during electrolytic charging of steel, Corrosion Science 87
  (2014) 239--258.

\bibitem{koren_experimental_2023}
E.~Koren, C.~M.~H.~Hagen, D.~Wang, X.~Lu, R.~Johnsen, J.~Yamabe, Experimental
  comparison of gaseous and electrochemical hydrogen charging in {X65} pipeline
  steel using the permeation technique, Corrosion Science 215 (2023) 111025.

\bibitem{wang_study_2023}
C.~Wang, J.~Zhang, C.~Liu, Q.~Hu, R.~Zhang, X.~Xu, H.~Yang, Y.~Ning, Y.~Li,
  Study on hydrogen embrittlement susceptibility of {X80} steel through in-situ
  gaseous hydrogen permeation and slow strain rate tensile tests, International
  Journal of Hydrogen Energy 48~(1) (2023) 243--256.

\bibitem{gao_synergic_2024}
R.~Gao, B.~Xing, C.~Yang, X.~Jiang, J.~Shang, Z.~Hua, Synergic effects of
  temperature and pressure on the hydrogen diffusion and dissolution behaviour
  of {X80} pipeline steel, Corrosion Science 240 (2024) 112468.

\bibitem{fan_hydrogen_2025}
X.~Fan, Y.~F. Cheng, Hydrogen pipelines and embrittlement in gaseous
  environments: {An} up-to-date review, Applied Energy 387 (2025) 125636.

\bibitem{okayasu2021examination}
M.~Okayasu, M.~Sato, Examination of hydrogen diffusivity in carbon steels using
  a newly developed hydrogen permeation system, Experimental Mechanics 61~(9)
  (2021) 1443--1453.

\bibitem{protopopoff2002surface}
E.~Protopopoff, P.~Marcus, Surface effects on hydrogen entry into metals, in:
  Corrosion mechanisms in theory and practice, CRC Press, 2002, pp. 62--105.

\bibitem{zhang_gaseous_2025}
R.~Zhang, C.~Wang, C.~Liu, H.~Zhang, M.~Zhu, Y.~Song, T.~Zhang, Y.~Li, Gaseous
  hydrogen permeation of pipeline steels: {A} focused review, Renewable and
  Sustainable Energy Reviews 211 (2025) 115304.

\bibitem{rothig_gaseous_2025}
M.~Röthig, J.~Hoschke, M.~F.~W. Chowdhury, C.~V. Tapia-Bastidas, J.~Venezuela,
  E.~Gray, T.~Depover, K.~Verbeken, M.~B. Djukic, A.~Atrens, Gaseous hydrogen
  permeation in {X65} {D} pipeline steel and a preliminary evaluation of the
  influence of oxygen, International Journal of Hydrogen Energy 158 (2025)
  150459.

\bibitem{fujiwara_high-pressure_2020}
H.~Fujiwara, H.~Ono, K.~Onoue, S.~Nishimura, High-pressure gaseous hydrogen
  permeation test method -property of polymeric materials for high-pressure
  hydrogen devices (1)-, International Journal of Hydrogen Energy 45~(53)
  (2020) 29082--29094.

\bibitem{huang_effect_2020}
H.~Huang, J.~Zheng, S.~Ding, W.~Wang, H.~Zhang, Effect of natural oxide film on
  the deuterium permeation behavior of 430 stainless steel, Fusion Engineering
  and Design 152 (2020) 111469.

\bibitem{sand_versatile_2024}
P.~Sand, A.~Manhard, U.~von Toussaint, A versatile setup for hydrogen isotope
  permeation studies, Review of Scientific Instruments 95~(12) (2024) 123302.

\bibitem{wampler_surface-limited_1986}
W.~R. Wampler, Surface-limited release of deuterium from iron and the effect of
  surface oxygen, Applied Physics Letters 48~(6) (1986) 405--407.

\bibitem{wampler_hydrogen_1989}
W.~R. Wampler, Hydrogen permeation into iron, Journal of Applied Physics
  65~(10) (1989) 4040--4044.

\bibitem{AnodicIronOxide1992a}
P.~Bruzzoni, R.~Garavaglia, Anodic iron oxide films and their effect on the
  hydrogen permeation through steel, Corrosion Science 33~(11) (1992)
  1797--1807.

\bibitem{roosendaal_passivation_1999}
S.~J. Roosendaal, Passivation mechanisms in the initial oxidation of iron by
  oxygen and water vapor: {Sander} {Jurgen} {Roosendaal}, Ph.D. thesis,
  Uitgever niet vastgesteld, Plaats van uitgave niet vastgesteld, oCLC:
  1012263133 (1999).

\bibitem{grosvenor_studies_2004}
A.~P. Grosvenor, B.~A. Kobe, N.~S. McIntyre, Studies of the oxidation of iron
  by air after being exposed to water vapour using angle‐resolved x‐ray
  photoelectron spectroscopy and {QUASES}$^{\textrm{™}}$, Surface and
  Interface Analysis 36~(13) (2004) 1637--1641.

\bibitem{somerday2013elucidating}
B.~P. Somerday, P.~Sofronis, K.~A. Nibur, C.~San~Marchi, R.~Kirchheim,
  Elucidating the variables affecting accelerated fatigue crack growth of
  steels in hydrogen gas with low oxygen concentrations, Acta Materialia
  61~(16) (2013) 6153--6170.

\bibitem{komoda2014effect}
R.~Komoda, M.~Kubota, Y.~Kondo, J.~Furtado, Effect of oxygen addition on
  fretting fatigue strength in hydrogen of jis sus304 stainless steel,
  Tribology International 76 (2014) 92--99.

\bibitem{nibur2024non}
K.~Nibur, Non-conservative fracture toughness measurements due to unintentional
  trace oxygen impurities in hydrogen gas, in: Pressure Vessels and Piping
  Conference, Vol. 88506, American Society of Mechanical Engineers, 2024, p.
  V004T06A016.

\bibitem{volklDiffusionHydrogenMetals1978}
J.~Völkl, G.~Alefeld, Diffusion of hydrogen in metals, in: S.~Amelinckx, V.~P.
  Chebotayev, R.~Gomer, H.~Ibach, V.~S. Letokhov, H.~K.~V. Lotsch, H.~J.
  Queisser, F.~P. Schäfer, A.~Seeger, K.~Shimoda, T.~Tamir, W.~T. Welford,
  H.~P.~J. Wijn, G.~Alefeld, J.~Völkl (Eds.), Hydrogen in {Metals} {I},
  Vol.~28, Springer Berlin Heidelberg, Berlin, Heidelberg, 1978, pp. 321--348,
  series Title: Topics in Applied Physics.

\bibitem{kiuchi_solubility_1983}
K.~Kiuchi, R.~B. McLellan, The solubility and diffusivity of hydrogen in
  well-annealed and deformed iron, Acta Metallurgica 31~(7) (1983) 961--984.

\bibitem{cupertino2023hydrogen}
L.~Cupertino-Malheiros, A.~Oudriss, F.~Th{\'e}bault, M.~Piette, X.~Feaugas,
  Hydrogen diffusion and trapping in low-alloy tempered martensitic steels,
  Metallurgical and Materials Transactions A 54~(4) (2023) 1159--1173.

\bibitem{san2007permeability}
C.~San~Marchi, B.~P. Somerday, S.~L. Robinson, Permeability, solubility and
  diffusivity of hydrogen isotopes in stainless steels at high gas pressures,
  International Journal of Hydrogen Energy 32~(1) (2007) 100--116.

\bibitem{oriani1970diffusion}
R.~A. Oriani, The diffusion and trapping of hydrogen in steel, Acta
  metallurgica 18~(1) (1970) 147--157.

\bibitem{turnbull2015perspectives}
A.~Turnbull, Perspectives on hydrogen uptake, diffusion and trapping,
  International Journal of Hydrogen Energy 40~(47) (2015) 16961--16970.

\bibitem{diaz2025comsol}
A.~D{\'\i}az, J.~M. Alegre, I.~I. Cuesta, E.~Mart{\'\i}nez-Pa{\~n}eda, A comsol
  framework for predicting hydrogen embrittlement, part i: Coupled hydrogen
  transport, Engineering Fracture Mechanics 319 (2025) 111007.

\bibitem{young_measurement_2020}
K.~T. Young, T.~M. Krentz, A.~L. d’Entremont, E.~M. Vogel, D.~A. Hitchcock,
  Measurement of gas-concentration-driven permeation for the examination of
  permeability, solubility, and diffusivity in varying materials, Review of
  Scientific Instruments 91~(10) (2020) 105105.

\bibitem{chalfoun_hydrogen_2022}
D.~R. Chalfoun, M.~A. Kappes, P.~Bruzzoni, M.~Iannuzzi, Hydrogen solubility,
  diffusivity, and trapping in quenched and tempered {Ni}-containing steels,
  International Journal of Hydrogen Energy 47~(5) (2022) 3141--3156.

\bibitem{bryan_diffusivity_1963}
W.~L. Bryan, B.~F. Dodge, Diffusivity of hydrogen in pure iron, AIChE Journal
  9~(2) (1963) 223--228.

\bibitem{mendibide_effect_2024}
C.~Mendibide, F.~Vucko, M.~Martinez, G.~Joshi, J.~Kittel, Effect of degraded
  environmental conditions on the service behavior of a {X65} pipeline steel
  not designed for hydrogen transport, International Journal of Hydrogen Energy
  52 (2024) 1019--1032.

\bibitem{miller_permeation_1975}
R.~F. Miller, J.~B. Hudson, G.~S. Ansell, Permeation of hydrogen through alpha
  iron, Metallurgical Transactions A 6~(1) (1975) 117--121.

\bibitem{gilroy_oxidation_1965}
D.~Gilroy, J.~E.~O. Mayne, The oxidation of iron at room temperature, Corrosion
  Science 5~(1) (1965) 55--58.

\bibitem{krugerRoomTemperatureOxidation1964}
J.~Kruger, H.~T. Yolken, Room {{Temperature Oxidation}} of {{Iron}} at {{Low
  Pressures}}, CORROSION 20~(1)  29t--33t.

\bibitem{grosvenorUseQUASESXPS2004}
A.~P. Grosvenor, B.~A. Kobe, N.~S. McIntyre, S.~Tougaard, W.~N. Lennard, Use of
  {QUASES}™/{XPS} measurements to determine the oxide composition and
  thickness on an iron substrate, Surface and Interface Analysis 36~(7) (2004)
  632--639.

\bibitem{bhargava_characterization_2007}
G.~Bhargava, I.~Gouzman, C.~Chun, T.~Ramanarayanan, S.~Bernasek,
  Characterization of the “native” surface thin film on pure
  polycrystalline iron: {A} high resolution {XPS} and {TEM} study, Applied
  Surface Science 253~(9) (2007) 4322--4329.

\bibitem{Nelson1973}
H.~G. Nelson, J.~E. Stein, Gas-phase hydrogen permeation through alpha iron,
  4130 steel, and 304 stainless steel from less than 100°c to near 600°c,
  Tech. rep., {National Aeronautics and Space Administration, Moffett Field,
  Calif. (USA). Ames Research Center}, United States (Apr. 1973).

\bibitem{girottoEffectPhysicochemicalProperties2023}
C.~P. Girotto, R.~P. Nippes, P.~D. Macruz, A.~D. Gomes, M.~De~Souza, M.~T.
  Rodriguez, Effect of physicochemical properties on the performance of
  palladium-based composite membranes: {{A}} review, Journal of Materials
  Research 38~(22) (2023) 4868--4891.

\bibitem{manolatos_necessity_1995}
P.~Manolatos, M.~Jerome, J.~Galland, Necessity of a palladium coating to ensure
  hydrogen oxidation during electrochemical permeation measurements on iron,
  Electrochimica Acta 40~(7) (1995) 867--871.

\bibitem{houben_comparison_2019}
A.~Houben, J.~Engels, M.~Rasiński, C.~Linsmeier, Comparison of the hydrogen
  permeation through fusion relevant steels and the influence of oxidized and
  rough surfaces, Nuclear Materials and Energy 19 (2019) 55--58.

\bibitem{vucko_hydrogen_2022}
F.~Vucko, S.~Ootsuka, S.~Rioual, E.~Diler, A.~Nazarov, D.~Thierry, Hydrogen
  detection in high strength dual phase steel using scanning {Kelvin} probe
  technique and {XPS} analyses, Corrosion Science 197 (2022) 110072.

\bibitem{linstromNISTChemistryWebBook1997}
P.~Linstrom, {{NIST Chemistry WebBook}}, {{NIST Standard Reference Database}}
  69.

\bibitem{thomas_c_allison_nist-janaf_2013}
T.~C. Allison, {NIST}-{JANAF} {Thermochemical} {Tables} - {SRD} 13 (Jan. 2013).

\bibitem{zafra_relative_2023}
A.~Zafra, Z.~Harris, E.~Korec, E.~Martínez-Pañeda, On the relative efficacy
  of electropermeation and isothermal desorption approaches for measuring
  hydrogen diffusivity, International Journal of Hydrogen Energy 48~(3) (2023)
  1218--1233.

\bibitem{fradet_thermochemical_2023}
Q.~Fradet, M.~Kurnatowska, U.~Riedel, Thermochemical reduction of iron oxide
  powders with hydrogen: {Review} of selected thermal analysis studies,
  Thermochimica Acta 726 (2023) 179552.

\bibitem{ma_influence_2021}
H.~C. Ma, D.~Zagidulin, M.~Goldman, D.~W. Shoesmith, Influence of iron oxides
  and calcareous deposits on the hydrogen permeation rate in {X65} steel in a
  simulated groundwater, International Journal of Hydrogen Energy 46~(9) (2021)
  6669--6679.

\bibitem{pineau_kinetics_2006}
A.~Pineau, N.~Kanari, I.~Gaballah, Kinetics of reduction of iron oxides by
  {H2}, Thermochimica Acta 447~(1) (2006) 89--100.

\bibitem{spreitzer_reduction_2019}
D.~Spreitzer, J.~Schenk, Reduction of {Iron} {Oxides} with {Hydrogen}—{A}
  {Review}, steel research international 90~(10) (2019) 1900108.

\bibitem{yarar_pd_2023}
M.~Yarar, A.~Bouziani, D.~Uner, Pd as a reduction promoter for {TiO2}: {Oxygen}
  and hydrogen transport at {2D} and {3D} {Pd} interfaces with {TiO2} monitored
  by {TPR}, \textit{operando} {1H} {NMR} and {CO} oxidation studies, Catalysis
  Communications 174 (2023) 106580.

\bibitem{oconnor_hydrogen_2020}
C.~R. O’Connor, M.~A. van Spronsen, T.~Egle, F.~Xu, H.~R. Kersell,
  J.~Oliver-Meseguer, M.~Karatok, M.~Salmeron, R.~J. Madix, C.~M. Friend,
  Hydrogen migration at restructuring palladium–silver oxide boundaries
  dramatically enhances reduction rate of silver oxide, Nature Communications
  11 (2020) 1844.

\bibitem{diaz_simulation_2020}
A.~Díaz, A.~Zafra, E.~Martínez-Pañeda, J.~Alegre, J.~Belzunce, I.~Cuesta,
  Simulation of hydrogen permeation through pure iron for trapping and surface
  phenomena characterisation, Theoretical and Applied Fracture Mechanics 110
  (2020) 102818.

\bibitem{kommlingInsightsLifetimePredictions2020}
A.~K{\"o}mmling, M.~Jaunich, M.~Goral, D.~Wolff, Insights for lifetime
  predictions of {{O-ring}} seals from five-year long-term aging tests, Polymer
  Degradation and Stability 179 (2020) 109278.

\bibitem{zhou_impact_2024}
C.~Zhou, H.~Zhou, L.~Zhang, The {Impact} of {Impurity} {Gases} on the
  {Hydrogen} {Embrittlement} {Behavior} of {Pipeline} {Steel} in
  {High}-{Pressure} {H2} {Environments}, Materials 17~(9) (2024) 2157.

\bibitem{garcia2024tds}
E.~Garc{\'\i}a-Mac{\'\i}as, Z.~D. Harris, E.~Mart{\'\i}nez-Pa{\~n}eda, Tds
  simulator: A matlab app to model temperature-programmed hydrogen desorption,
  International Journal of Hydrogen Energy 94 (2024) 510--524.

\bibitem{o.d.gonzalezPARTVICommunications1968}
{O. D. Gonzalez}, {PART} {VI} - {Communications} - {Permeation} of {Hydrogen}
  and {Deuterium} in {Alpha} {Iron}, The American Institute of Mining,
  Metallurgical, and Petroleum Engineers, 1968.

\bibitem{choi_diffusion_1970}
J.~Y. Choi, Diffusion of hydrogen in iron, Metallurgical Transactions 1~(4)
  (1970) 911--919.

\bibitem{SanMarchi2012}
C.~San~Marchi, B.~Somerday, {Technical Reference for Hydrogen Compatibility of
  Materials}, Sandia Report (2012).

\bibitem{taharaMeasurementsPermeationHydrogen1985}
A.~Tahara, Y.~Hayashi, Measurements of {{Permeation}} of {{Hydrogen Isotopes}}
  through {$\alpha$}-{{Iron}} by {{Pressure Modulation}} and {{Ion
  Bombarding}}, Transactions of the Japan Institute of Metals 26~(12) (1985)
  869--875.

\bibitem{gadgeel_gas-phase_1979}
V.~L. Gadgeel, D.~L. Johnson, Gas-phase hydrogen permeation and diffusion in
  carbon steels as a function of carbon content from 500 to 900 {K}, Journal of
  Materials for Energy Systems 1~(2) (1979) 32--40.

\bibitem{quickHydrogenDeuteriumIron1978}
N.~R. Quick, H.~H. Johnson, Hydrogen and deuterium in iron,
  49--506{$^\circ$}{{C}}, Acta Metallurgica 26~(6) (1978) 903--907.

\bibitem{kumnickHydrogenDeuteriumIron1977}
A.~J. Kumnick, H.~H. Johnson, Hydrogen and deuterium in iron, 9–73°{C}, Acta
  Metallurgica 25~(8) (1977) 891--895.

\bibitem{johnsonDiffusivityHydrogenAlpha1960}
E.~W. Johnson, M.~L. Hill, The {{Diffusivity}} of {{Hydrogen}} in {{Alpha
  Iron}}, Transactions of the Metallurgical Society of AIME 218 (1960)
  1104--1111.

\bibitem{tanabe_hydrogen_1984}
T.~Tanabe, Y.~Yamanishi, S.~Imoto, Hydrogen {Transport} through {Highly}
  {Purified} {Iron}, Transactions of the Japan Institute of Metals 25~(1)
  (1984) 1--10.

\bibitem{hagi_diffusion_1979}
H.~Hagi, Y.~Hayashi, N.~Ohtani, Diffusion {Coefficient} of {Hydrogen} in {Pure}
  {Iron} between 230 and 300 {K}, Transactions of the Japan Institute of Metals
  20~(7) (1979) 349--357.

\bibitem{zhang_effects_2018}
T.~Zhang, W.~Zhao, Y.~Zhao, K.~Ouyang, Q.~Deng, Y.~Wang, W.~Jiang, Effects of
  surface oxide films on hydrogen permeation and susceptibility to
  embrittlement of {X80} steel under hydrogen atmosphere, International
  Journal of Hydrogen Energy 43~(6) (2018) 3353--3365.

\bibitem{wang_atom_2025}
S.~Wang, Z.~Wang, L.~Jiang, P.~Zhang, M.~Y. Tan, R.~K.~W. Marceau, Atom probe
  tomography investigation of the effects of {Fe}-oxide layers on hydrogen
  penetration in pipeline steel, International Journal of Hydrogen Energy 161
  (2025) 150696.

\bibitem{kiuchi_effect_1983}
K.~Kiuchi, R.~B. McLellan, The effect of surface contamination on the measured
  hydrogen solubilities of metals, Journal of the Less Common Metals 95~(2)
  (1983) 283--292.

\bibitem{martinez-paneda_generalised_2020}
E.~Martínez-Pañeda, A.~Díaz, L.~Wright, A.~Turnbull, Generalised boundary
  conditions for hydrogen transport at crack tips, Corrosion Science 173 (2020)
  108698.

\end{thebibliography}


\end{document}